\documentclass[pdflatex,sn-mathphys-num]{sn-jnl}% Math and Physical Sciences Numbered Reference Style
\usepackage{graphicx}%
\usepackage{multirow}%
\usepackage{amsmath,amssymb,amsfonts}%
\usepackage{amsthm}%
\usepackage{mathrsfs}%
\usepackage[title]{appendix}%
\usepackage{xcolor}%
\usepackage{textcomp}%
\usepackage{manyfoot}%
\usepackage{booktabs}%
\usepackage{algorithm}%
\usepackage{algorithmicx}%
\usepackage{algpseudocode}%
\usepackage{listings}%
\usepackage{tabularx}
\usepackage{ragged2e}
\usepackage{wasysym}
\usepackage{float}
\usepackage{wrapfig}
\usepackage{makecell}
\usepackage[dvipsnames]{xcolor}
\usepackage{newtxtext}
\usepackage{newtxmath}
\usepackage{natbib}
\usepackage{hyperref}
\usepackage{gensymb}

\theoremstyle{thmstyleone}%
\theoremstyle{thmstyletwo}%

\theoremstyle{thmstylethree}%

\begin{document}

\title[Near-wall sedimentation of short cylinders at low Reynolds numbers: Translational and angular velocity models and experimental validation]{Near-wall sedimentation of short cylinders at low Reynolds numbers: Translational and angular velocity models and experimental validation}


\author[1]{\fnm{Isabell} \sur{Noichl}}\email{isabell.noichl@mv.rptu.de}

\author*[1,2]{\fnm{Clarissa} \sur{Sch\"onecker}}\email{clarissa.schoenecker@h-da.de}

\affil*[1]{\orgname{RPTU University Kaiserslautern-Landau}, \orgaddress{\city{Kaiserslautern}, \postcode{D-67663}, \country{Germany}}}

\affil[2]{\orgname{Darmstadt University of Applied Sciences}, \orgaddress{\city{Darmstadt}, \postcode{D-64295}, \country{Germany}}}


\abstract{The sedimentation of short cylinders near rigid walls involves coupled translational and rotational motion, for which readily applicable kinematic descriptions remain limited. In this study, theoretical expressions for the translational and angular velocities were derived by combining slender-body theory, semi-empirical finite-aspect-ratio corrections, and hydrodynamic descriptions of fluid inertia and wall effects. The resulting expressions were assessed through time-resolved sedimentation experiments with rigid cylinders of mean aspect ratios $\bar{\chi}\approx3.8$--$5.8$ in a rectangular duct at particle Reynolds numbers of $Re_{\mathrm P}\approx O(10^{-1})$. The experiments revealed distinct near-wall trajectories, including non-contact glancing motion and contact-induced flipping. For glancing trajectories, the predicted translational and angular velocities reproduced the principal features of the measured kinematics, with good qualitative and, over substantial parts of the trajectories, quantitative agreement. The comparisons also identified different ranges of applicability for the investigated wall corrections and showed that wall-induced effects dominated the angular velocity under the conditions studied. The resulting, experimentally validated expressions provide readily applicable estimates of the instantaneous translational and angular velocities of short cylinders near rigid walls and a reference for investigating additional influences on their sedimentation dynamics, including lower Reynolds numbers and particle or wall elasticity.}




\maketitle

\section{Introduction}\label{sec1}

The sedimentation of cylindrical and rod-shaped particles is relevant to
a wide range of environmental, biological, and industrial processes.
Examples include the transport and deposition of natural and synthetic
fibres, fibrous microplastics, and rod-shaped biological particles. In
applications involving particle transport, the settling velocity is
frequently a quantity of primary practical interest because it influences
particle residence times, transport distances, and deposition. Accordingly,
recent studies continue to seek predictive descriptions of the settling
velocities of non-spherical particles. Hamidi \textit{et al.}
\cite{Hamidi.2026}, for example, experimentally investigated the settling
of straight and bent rods in a quiescent fluid and developed a
semi-empirical description of their terminal settling velocities,
highlighting the importance of particle morphology for transport
predictions.

In other applications, the motion of a particle relative to a nearby
surface is equally important. For instance, the approach of rod-shaped
biological particles toward a solid boundary raises questions about their
orientation, lateral motion, and rotational behaviour before surface
contact and possible attachment. Understanding these near-wall kinematics
is therefore relevant to the physical processes preceding biofilm
formation, although attachment itself also depends on biological and
physicochemical interactions. Unlike a spherical particle, a cylinder
exhibits direction-dependent hydrodynamic resistance. Its translational
motion depends on its orientation relative to the direction of motion,
while nearby boundaries can additionally couple translation and rotation.
Consequently, describing the motion of a freely sedimenting cylinder near
a wall requires consideration of the simultaneous evolution of its
orientation, wall distance, translational velocity, and angular velocity.

The hydrodynamics of cylindrical particles in unbounded fluids have been
investigated using analytical, numerical, and experimental approaches.
Classical slender-body theory provides descriptions of the translational
and rotational resistance of elongated particles in the creeping-flow
limit \cite{Batchelor.1970}. For short cylinders, however, finite-length
effects, including the influence of the cylinder ends, become increasingly
important. Kharrouba \textit{et al.} \cite{Kharrouba.2021} investigated
the hydrodynamic loads on finite-length cylinders over a range of aspect
ratios, inclinations, and Reynolds numbers. Pierson \textit{et al.}
\cite{Pierson.2021} subsequently examined the resistance torque on a
cylinder rotating about an axis perpendicular to its symmetry axis and
derived higher-order slender-body approximations. These studies provide
descriptions of hydrodynamic resistance relevant to the short cylinders
considered here.

Even in an unbounded fluid, weak fluid inertia can modify the
sedimentation of an inclined cylinder by generating a hydrodynamic
orienting torque. Khayat and Cox \cite{Khayat.1989} derived inertial
corrections to the hydrodynamic force and torque on slender bodies.
Cabrera-Booman \textit{et al.} \cite{Cabrera.2022} experimentally
investigated freely settling cylinders and compared their measured
motion with theoretical descriptions of fluid-inertia effects. These
studies demonstrate the importance of orientation and weak inertia in
cylinder sedimentation, even without a nearby boundary.

A rigid wall introduces further complexity because the hydrodynamic
resistance and induced rotation depend on the instantaneous
particle--wall configuration. De Mestre and Russel
\cite{deMestre.1975} developed a slender-body description of the
hydrodynamic force and torque on a cylinder translating near a plane
wall. Russel \textit{et al.} \cite{RusselWB.1977} subsequently combined
theoretical analysis and experiments to investigate slender rods
sedimenting near a vertical wall. Their results showed how wall-induced
rotation can change the lateral motion of an inclined rod and produce
distinct near-wall trajectories, including turns away from the wall
without contact. More recently, Koens and Montenegro-Johnson
\cite{Koens.2021} developed an asymptotic description of the local drag
on a slender rod parallel to a plane wall. Chun and Jung
\cite{Chun.2025} numerically investigated rod-like particles in
confined shear flows and demonstrated that wall proximity and the
imposed flow configuration can substantially modify their rotational
motion, including tumbling-associated pole-vaulting behaviour. Although
their imposed-shear configuration differs from freely sedimenting
cylinders in an otherwise quiescent fluid, these investigations
illustrate the sensitivity of rod kinematics to the surrounding
boundaries and flow conditions.

The near-wall sedimentation of a short cylinder is particularly
challenging because neither its orientation nor its distance from the
wall generally remains constant. Both quantities evolve during the
motion and influence the instantaneous translational and angular
velocities. Furthermore, while recent experimental work has sought
practically usable descriptions of terminal settling velocities
\cite{Hamidi.2026}, the evaluation of such descriptions at a steady
state does not establish their applicability throughout a
time-dependent near-wall trajectory. A comparison between measured
instantaneous velocities and quasi-steady model velocities along the
observed trajectory can therefore provide information about the
applicability and limitations of the underlying hydrodynamic
approximations.

The choice of particle Reynolds number is also motivated by previous
experiments on wall-bounded sphere sedimentation
\cite{Noichl.2026}. In that system, rigid spheres sedimenting at
$Re_{\mathrm P}\approx O(10^{-1})$ exhibited post-acceleration
near-wall motion consistent with classical wall-lift descriptions,
whereas experiments at $Re_{\mathrm P}\approx O(10^{-2})$ revealed
persistently unsteady sedimentation despite the small particle
Reynolds numbers. These observations do not establish a universal
transition Reynolds number or imply that cylinders exhibit the same
behaviour. They do, however, motivate the investigation of short,
rigid cylinders at $Re_{\mathrm P}\approx O(10^{-1})$ as a reference
for subsequent studies at smaller Reynolds numbers.

The present study experimentally investigates the sedimentation of
short, rigid cylinders with mean aspect ratios of approximately
$\bar{\chi}=3.8$--$5.8$ in a rectangular duct filled with a highly
viscous, initially quiescent liquid. The cylinders are released with
negligible initial translational velocity, either approximately
axially aligned near the centre of the duct or at different initial
inclinations and distances from a lateral rigid wall. Time-resolved
image measurements are used to determine their complete
trajectories, including the evolution of particle orientation, wall
distance, translational velocity, and angular velocity. The
centre-duct experiments first provide a reference for assessing
predicted sedimentation velocities. The near-wall experiments then
resolve the coupled translational and angular kinematics, including
glancing trajectories and contact-induced flipping behaviour.

To describe the measured kinematics, existing expressions for the
hydrodynamic resistance of finite-length cylinders and for
wall-induced motion are combined into models that provide
translational and angular velocities as functions of the
instantaneous particle configuration. The theoretical velocities
are evaluated using the experimentally measured orientation and
wall distance and compared with the corresponding measured
velocities throughout the evaluated trajectories. The objective is
therefore to assess the applicability of quasi-steady velocity
descriptions to the experimentally observed, time-dependent
near-wall motion, rather than to predict complete trajectories
from prescribed initial conditions. Expressing the models directly
in terms of kinematic quantities provides a comparatively
accessible description for applications in which particle
velocities, orientations, and surface-approach behaviour are of
primary interest.

The remainder of this paper is organized as follows. Section~2
introduces the theoretical descriptions used to estimate the
translational and angular velocities of freely sedimenting
cylinders, including the effects of finite aspect ratio, weak
fluid inertia, and confinement by a rigid wall. Section~3
describes the experimental setup, the cylindrical model particles,
and the image-based evaluation of their motion. Section~4 presents
and discusses the experimental results. First, the sedimentation
of approximately axially aligned cylinders near the centre of the
rectangular duct is examined. The subsequent analysis focuses on
the near-wall translational and angular kinematics of initially
inclined cylinders, comparing the instantaneous measured
velocities with the corresponding model estimates along the
experimentally observed trajectories. Finally, Section~5
summarizes the main findings, discusses the applicability and
limitations of the kinematic descriptions, and outlines
directions for further investigation.

\section{Theory}
\label{sec2}

\subsection{Translational velocity}
\label{sec2_1}

A rigid circular cylinder of radius \(r\), total length \(2l\), volume
\(V_{\mathrm{P}}=2\pi r^2l\), and density \(\rho_{\mathrm{P}}\) sediments
with velocity magnitude \(U=\lvert\boldsymbol{U}\rvert\) in an otherwise
quiescent fluid of density \(\rho\) and dynamic viscosity \(\eta\).
The density ratio approaches unity from above,
\(\gamma=\rho_{\mathrm{P}}/\rho\rightarrow1^{+}\), such that the density
difference between the particle and the surrounding fluid is small. The
cylinder aspect ratio is defined as \(\chi=l/r\).

The cylinder is released from rest at \(t=0\), while the surrounding fluid
is initially quiescent, \(\lvert\boldsymbol{u}_{\mathrm{f}}\rvert=0\).
The translational particle Reynolds number used throughout most of the
present formulation is defined as
\(Re_{\mathrm{P}}=\rho Ul/\eta\).

\begin{figure}
    \centering
    \includegraphics[width=0.7\textwidth]{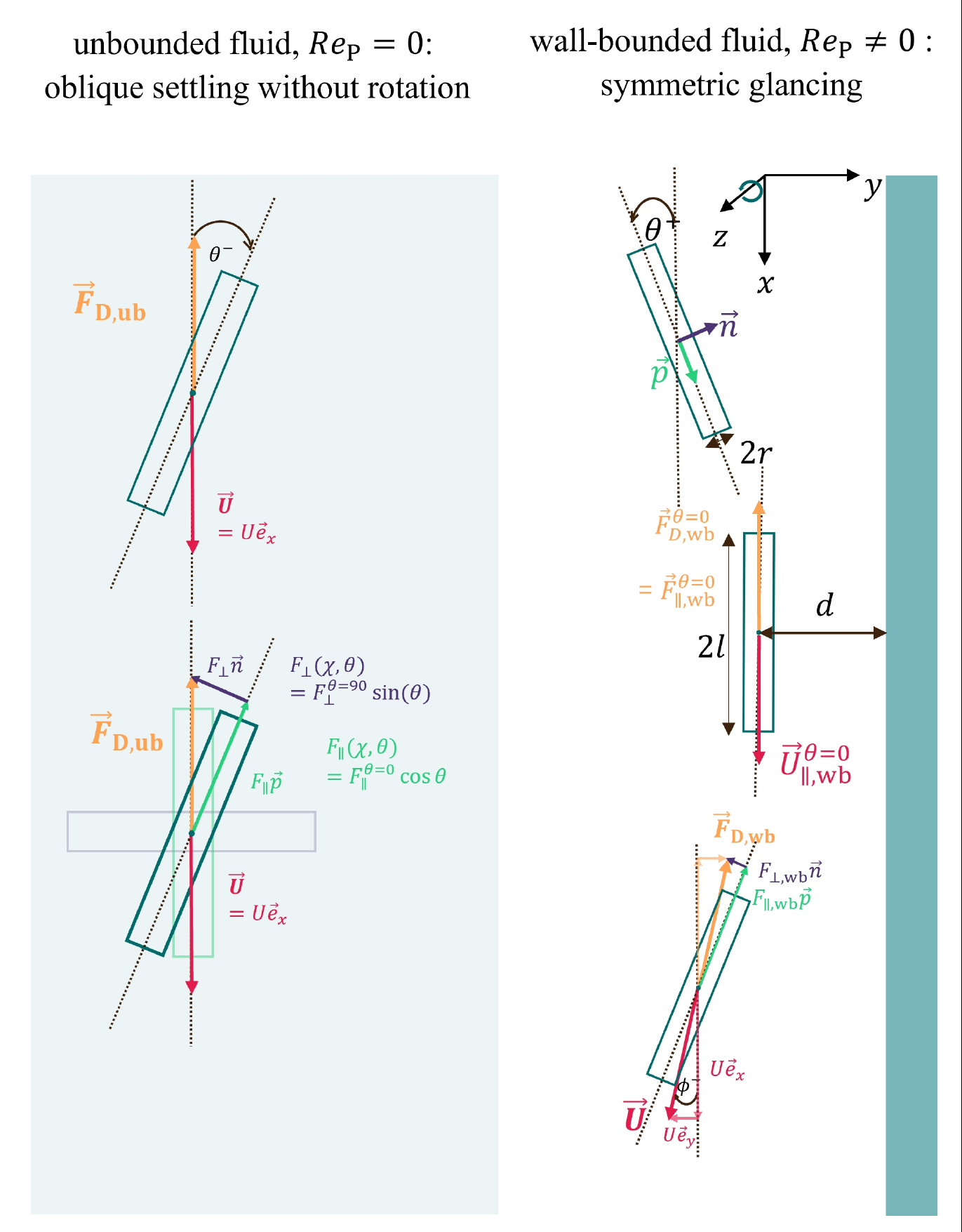}
    \caption{Definitions of the spatial and angular quantities and schematic
    force decompositions. Left: oblique sedimentation of a cylinder in an
    unbounded fluid in the creeping-flow regime. Right: translation and
    rotation of a cylinder near a plane wall at finite particle Reynolds
    number.}
    \label{Fig1}
\end{figure}

The spatial and angular quantities used throughout the following are
defined in Fig.~\ref{Fig1}. In particular, the orientation angle
\(\theta\) denotes the angle between the cylinder axis and the direction
of gravity, whereas the trajectory angle \(\phi\) denotes the angle
between the instantaneous translational velocity vector and the direction
of gravity. In the wall-bounded configuration, \(d\) denotes the distance
between the cylinder center and the plane wall.

For axial sedimentation, \(\theta=0^\circ\), the steady hydrodynamic drag
is balanced by the net gravitational force acting on the cylinder,
yielding
\begin{equation}
\label{eq1}
\left(\rho_{\mathrm{P}}-\rho\right)V_{\mathrm{P}}g
=
F_{\mathrm{D,ub}}^{\theta=0^\circ,\,Re_{\mathrm{P}}\approx0}.
\end{equation}
For transverse sedimentation, \(\theta=90^\circ\), the corresponding
force balance becomes
\begin{equation}
\label{eq2}
\left(\rho_{\mathrm{P}}-\rho\right)V_{\mathrm{P}}g
=
F_{\mathrm{D,ub}}^{\theta=90^\circ,\,Re_{\mathrm{P}}\approx0}.
\end{equation}

Kharrouba \textit{et al.}~\cite{Kharrouba.2021} developed semi-empirical
expressions for the drag force on cylinders with aspect ratios in the
range \(2\leq\chi\leq10\). Their formulation is based on the
slender-body approximations of Batchelor~\cite{Batchelor.1970},
Cox~\cite{Cox.1970}, and Keller and Rubinow~
\cite{JosephB.KellerandSolI.Rubinow.}. Higher-order corrections and an
empirical finite-length contribution obtained from numerical simulations
were incorporated to account for finite-aspect-ratio effects, which
become increasingly important for \(\chi\lesssim10\).

The axial drag in an unbounded fluid in the creeping-flow limit is
approximated as
\begin{equation}
\label{eq3}
\begin{aligned}
F_{\mathrm{D,ub}}^{\theta=0^\circ,\,Re_{\mathrm{P}}\approx0}
={}&
4\pi\eta lU \Bigg[
\frac{a_{\mathrm{x}}^{(1)}}{\ln(2\chi)}
+\frac{a_{\mathrm{x}}^{(2)}}{\left[\ln(2\chi)\right]^2}
+\frac{a_{\mathrm{x}}^{(3)}}{\left[\ln(2\chi)\right]^3}
\\
&\qquad
+\frac{a_{\mathrm{x}}^{(4)}}{\left[\ln(2\chi)\right]^4}
-\frac{2.4}
{\chi^{2/3}\left(\chi-\frac{1}{2}\right)^{1.75}}
\Bigg],
\end{aligned}
\end{equation}
where \(a_{\mathrm{x}}^{(1)}=1\),
\(a_{\mathrm{x}}^{(2)}\approx0.80685\),
\(a_{\mathrm{x}}^{(3)}\approx0.82854\), and
\(a_{\mathrm{x}}^{(4)}\approx1.45243\).

The corresponding transverse drag is
\begin{equation}
\label{eq4}
\begin{aligned}
F_{\mathrm{D,ub}}^{\theta=90^\circ,\,Re_{\mathrm{P}}\approx0}
={}&
8\pi\eta lU \Bigg[
\frac{a_{\mathrm{y}}^{(1)}}{\ln(2\chi)}
+\frac{a_{\mathrm{y}}^{(2)}}{\left[\ln(2\chi)\right]^2}
\\
&\qquad
+\left(1-e^{-c_1\chi^{c_2}}\right)
\left(
\frac{a_{\mathrm{y}}^{(3)}}{\left[\ln(2\chi)\right]^3}
+\frac{a_{\mathrm{y}}^{(4)}}{\left[\ln(2\chi)\right]^4}
\right)
\Bigg],
\end{aligned}
\end{equation}
with \(a_{\mathrm{y}}^{(1)}=1\),
\(a_{\mathrm{y}}^{(2)}\approx-0.19315\),
\(a_{\mathrm{y}}^{(3)}\approx0.21484\),
\(a_{\mathrm{y}}^{(4)}\approx0.38735\),
\(c_1=0.01\), and \(c_2=2.5\).

Based on the theory of Khayat and Cox~\cite{Khayat.1989}, Kharrouba
\textit{et al.}~\cite{Kharrouba.2021} further proposed semi-empirical
expressions that account for weak fluid-inertia effects. In the regime
\(2Re_{\mathrm{P}}=O(1)\), the inertia-corrected axial drag is
\begin{equation}
\label{eq5}
\begin{aligned}
F_{\mathrm{D,ub}}^{\theta=0^\circ,\,2Re_{\mathrm{P}}=O(1)}
={}&
4\pi\eta lU \Bigg[
\frac{a_{\mathrm{x}}^{(1)}}{\ln(2\chi)}
+\frac{a_{\mathrm{x}}^{(2)}+f_{\parallel}}
{\left[\ln(2\chi)\right]^2}
+\frac{a_{\mathrm{x}}^{(3)}+f_3f_{\parallel}}
{\left[\ln(2\chi)\right]^3}
\\
&\qquad
+\frac{a_{\mathrm{x}}^{(4)}+f_4f_{\parallel}}
{\left[\ln(2\chi)\right]^4}
-\frac{2.4}
{\chi^{2/3}\left(\chi-\frac{1}{2}\right)^{1.75}}
\Bigg],
\end{aligned}
\end{equation}
where \(f_{\parallel}\approx Re_{\mathrm{P}}/4\),
\(f_3=(2Re_{\mathrm{P}})^{0.07\chi^{0.5}}\), and
\(f_4=(2Re_{\mathrm{P}})^{0.03\chi^{0.9}}\).

For transverse motion, only the second-order term is corrected, yielding
\begin{equation}
\label{eq6}
\begin{aligned}
F_{\mathrm{D,ub}}^{\theta=90^\circ,\,2Re_{\mathrm{P}}=O(1)}
={}&
8\pi\eta lU \Bigg[
\frac{a_{\mathrm{y}}^{(1)}}{\ln(2\chi)}
+\frac{a_{\mathrm{y}}^{(2)}+f_{\perp}}
{\left[\ln(2\chi)\right]^2}
\\
&\qquad
+\left(1-e^{-c_1\chi^{c_2}}\right)
\left(
\frac{a_{\mathrm{y}}^{(3)}}{\left[\ln(2\chi)\right]^3}
+\frac{a_{\mathrm{y}}^{(4)}}{\left[\ln(2\chi)\right]^4}
\right)
\Bigg].
\end{aligned}
\end{equation}
In the limit \(2Re_{\mathrm{P}}\rightarrow0\),
\(f_{\perp}\rightarrow Re_{\mathrm{P}}/2\). For larger Reynolds numbers,
the complete correlation for \(f_{\perp}\) is given in Eq.~(C3) of
Kharrouba \textit{et al.}~\cite{Kharrouba.2021}.

At this point, the Reynolds-number notation requires some clarification.
The present work primarily uses the half-length-based particle Reynolds
number \(Re_{\mathrm{P}}=\rho Ul/\eta\), whereas Kharrouba
\textit{et al.} also express their finite-inertia correlations in terms
of a Reynolds number based on the cylinder diameter \(2r\). The two
definitions are related by
\(Re=2\rho Ur/\eta=2Re_{\mathrm{P}}r/l=2Re_{\mathrm{P}}/\chi\).
The factor \(2Re_{\mathrm{P}}\) appearing explicitly in Eqs.~\eqref{eq5}
and \eqref{eq6} is retained in the form used by Kharrouba
\textit{et al.}, whereas the diameter-based Reynolds number \(Re\) is
employed below where their correlations are stated in this form. This
distinction is maintained throughout to avoid mixing the different
characteristic length scales.

For an inclined cylinder in the Stokes regime, the hydrodynamic force can
be decomposed into components parallel and perpendicular to the cylinder
axis. For a torque-free cylinder sedimenting obliquely in an unbounded
fluid, these components are
\begin{equation}
\label{eq7}
F_{\parallel,\mathrm{ub}}^{Re_{\mathrm{P}}=0}(\chi,\theta)
=
F_{\mathrm{D,ub}}^{\theta=0^\circ,\,
Re_{\mathrm{P}}\approx0}\cos\theta
\end{equation}
and
\begin{equation}
\label{eq8}
F_{\perp,\mathrm{ub}}^{Re_{\mathrm{P}}=0}(\chi,\theta)
=
F_{\mathrm{D,ub}}^{\theta=90^\circ,\,
Re_{\mathrm{P}}\approx0}\sin\theta.
\end{equation}

Even in the creeping-flow limit, an inclined cylinder generally does not
sediment exactly parallel to gravity because its hydrodynamic resistance
is anisotropic. Consequently, the translational velocity is not, in
general, collinear with the gravitational force. What vanishes in this
limit is the inertia-induced torque: in an unbounded Stokes flow, the
orientation remains constant in time, while the anisotropic translational
resistance allows an oblique settling trajectory. The left-hand sketch
in Fig.~\ref{Fig1} therefore represents a constant-orientation
creeping-flow configuration rather than a general absence of transverse
particle motion.

At finite Reynolds number and particularly in the presence of a wall,
fluid inertia gives rise to hydrodynamic torque and modifies the
translational motion. A freely sedimenting cylinder therefore does not
generally retain a constant orientation angle. Nevertheless, for small
Reynolds numbers and moderate orientation angles, the corresponding
projection approach remains useful when the inertia-corrected drag is
employed.

When lateral motion occurs, the translational velocity is characterized
by the trajectory angle
\(\phi=\arctan(U_{\mathrm{P,y}}/U_{\mathrm{P,x}})\). Accordingly,
\(U=\sqrt{U_{\mathrm{P,x}}^2+U_{\mathrm{P,y}}^2}
=U_{\mathrm{P,x}}/\cos\phi=U_{\mathrm{P,x}}\sec\phi\).

Similarly, the magnitude of the total hydrodynamic drag follows from its
components parallel and perpendicular to the cylinder axis,
\(F_{\mathrm{D}}=
\sqrt{F_{\parallel}^2(\theta)+F_{\perp}^2(\theta)}\).
For small orientation and trajectory angles,
\(\theta,\phi\rightarrow0^\circ\), the streamwise component of the total
drag can be approximated by the axial drag of an aligned cylinder,
\(F_{\mathrm{D}}\cos\phi\approx F_{\parallel}^{\theta=0^\circ}\).
The total drag is therefore approximated as
\begin{equation}
\label{eq9}
\begin{aligned}
F_{\mathrm{D}}(\chi,\theta,\phi)
&\approx
\frac{
F_{\parallel}^{\theta=0^\circ,\,
2Re_{\mathrm{P}}=O(1)}
}{
\cos\phi
}
\\
&=
F_{\parallel}^{\theta=0^\circ,\,
2Re_{\mathrm{P}}=O(1)}
\sec\phi.
\end{aligned}
\end{equation}

Brenner~\cite{Brenner.1962} developed a general theory according to which
the leading-order wall correction for a particle of arbitrary
cross-section moving along a principal axis of hydrodynamic resistance
can be related to its drag in an unbounded fluid. The theory requires
the corresponding wall correction for a sphere to be known and assumes
that the characteristic particle dimensions are small compared with the
distance from the boundary.

For a particle with characteristic length \(l^\ast\) located at a
characteristic distance \(d^\ast\) from a boundary, Brenner's wall
correction may be expressed as
\begin{equation}
\label{eq10}
\frac{F_{\mathrm{D,wb}}^\ast}{F_{\mathrm{D,ub}}^\ast}
=
\frac{1}{
1
-C_1F_{\mathrm{D,ub}}^\ast
\left(\frac{l^\ast}{d^\ast}\right)
+O\left[
\left(\frac{l^\ast}{d^\ast}\right)^3
\right]
}.
\end{equation}
Here,
\(F_{\mathrm{D,wb}}^\ast=
F_{\mathrm{D,wb}}/(2\pi\eta l^\ast U_{\mathrm{th,wb}})\) and
\(F_{\mathrm{D,ub}}^\ast=
F_{\mathrm{D,ub}}/(2\pi\eta l^\ast U_{\mathrm{th,ub}})\)
denote the dimensionless wall-bounded and unbounded drag coefficients,
respectively.

Ui \textit{et al.}~\cite{Ui.1984} investigated rods of different aspect
ratios sedimenting along the centerline of circular tubes of diameter
\(D\). By varying the tube diameter, measuring the steady sedimentation
velocity, and determining the dimensionless drag from the force balance,
they extrapolated the unbounded-fluid drag to \(l/(D/2)=0\). They found
that the linear relation
\begin{equation}
\label{eq11}
\frac{F_{\mathrm{D,wb}}^\ast}{F_{\mathrm{D,ub}}^\ast}
=
1
+C_1F_{\mathrm{D,ub}}^\ast
\left(\frac{l^\ast}{d^\ast}\right)
\end{equation}
provided a more suitable description of their measurements than the
corresponding expansion of Brenner's relation. The third-order
correction is omitted in Eq.~\eqref{eq11}, since the extrapolation by Ui
\textit{et al.} was restricted to the linear regime.

Happel and Bart~\cite{Happel.1974} derived the wall-bounded drag on a
sphere sedimenting along the centerline of a channel with a square
cross-section. For a square duct, the corresponding wall coefficient is
\(C_1=1.903266\), and the characteristic wall distance is the duct
half-width \(W/2\).

Using the first-order Brenner correction, hereafter denoted model (1),
the steady velocity of an axially sedimenting cylinder along the
centerline of a square duct is approximated by
\begin{equation}
\label{eq12}
\begin{aligned}
\left.
U_{\mathrm{th,wb}}^{\theta=0^\circ,\,Re_{\mathrm{P}}\approx0}
\right|_{\boxdot}^{(1)}
\approx{}&
\frac{
\left(\rho_{\mathrm{P}}-\rho\right)r^2g
}{
2\eta
F_{\mathrm{D,ub}}^{\ast,\theta=0^\circ,\,
Re_{\mathrm{P}}\approx0}
}
\\
&\times
\Bigg[
1
-1.903266
F_{\mathrm{D,ub}}^{\ast,\theta=0^\circ,\,
Re_{\mathrm{P}}\approx0}
\left(\frac{2l}{W}\right)
\Bigg].
\end{aligned}
\end{equation}

Using the first-order relation proposed by Ui \textit{et al.}, hereafter
denoted model (2), the corresponding velocity becomes
\begin{equation}
\label{eq13}
\begin{aligned}
\left.
U_{\mathrm{th,wb}}^{\theta=0^\circ,\,Re_{\mathrm{P}}\approx0}
\right|_{\boxdot}^{(2)}
\approx{}&
\frac{
\left(\rho_{\mathrm{P}}-\rho\right)r^2g
}{
2\eta
F_{\mathrm{D,ub}}^{\ast,\theta=0^\circ,\,
Re_{\mathrm{P}}\approx0}
}
\\
&\times
\Bigg[
1
+1.903266
F_{\mathrm{D,ub}}^{\ast,\theta=0^\circ,\,
Re_{\mathrm{P}}\approx0}
\left(\frac{2l}{W}\right)
\Bigg]^{-1}.
\end{aligned}
\end{equation}

In our previous study~\cite{Noichl.2026}, we showed that the drag on a
rigid sphere sedimenting asymmetrically in a rectangular duct at
\(Re_{\mathrm{P}}=O(10^{-1})\) is governed predominantly by the nearest
plane wall, whereas the contributions of the remaining walls are
comparatively small. In the near-wall region, at \(d/R\approx2\), the
sphere velocity was accurately approximated using Faxén's wall
correction~\cite{Faxen.1922b}.

Following Brenner's approach, the wall-induced drag correction for an
axially sedimenting cylinder near a plane wall can analogously be
constructed using the Faxén expansion. The resulting wall-bounded
velocity at a cylinder--wall distance \(d\) is
\begin{equation}
\label{eq14}
\begin{aligned}
\left.
U_{\mathrm{th,wb}}^{\theta=0^\circ,\,Re_{\mathrm{P}}\approx0}
\right|_{\mid\shortmid}^{(1)}
\approx{}&
\frac{
\left(\rho_{\mathrm{P}}-\rho\right)r^2g
}{
2\eta
F_{\mathrm{D,ub}}^{\ast,\theta=0^\circ,\,
Re_{\mathrm{P}}\approx0}
}
\\
&\times
\Bigg[
1
-\frac{9}{16}
F_{\mathrm{D,ub}}^{\ast,\theta=0^\circ,\,
Re_{\mathrm{P}}\approx0}
\left(\frac{l}{d}\right)
\\
&\qquad
+\frac{1}{8}
F_{\mathrm{D,ub}}^{\ast,\theta=0^\circ,\,
Re_{\mathrm{P}}\approx0}
\left(\frac{l}{d}\right)^3
\Bigg].
\end{aligned}
\end{equation}
Equation~\eqref{eq14} is based on Brenner's formulation and includes the
first- and third-order wall corrections. Based on the relation proposed
by Ui \textit{et al.}, and retaining only the first-order Faxén
correction, model (2) becomes
\begin{equation}
\label{eq15}
\begin{aligned}
\left.
U_{\mathrm{th,wb}}^{\theta=0^\circ,\,Re_{\mathrm{P}}\approx0}
\right|_{\mid\shortmid}^{(2)}
\approx{}&
\frac{
\left(\rho_{\mathrm{P}}-\rho\right)r^2g
}{
2\eta
F_{\mathrm{D,ub}}^{\ast,\theta=0^\circ,\,
Re_{\mathrm{P}}\approx0}
}
\\
&\times
\Bigg[
1
+\frac{9}{16}
F_{\mathrm{D,ub}}^{\ast,\theta=0^\circ,\,
Re_{\mathrm{P}}\approx0}
\left(\frac{l}{d}\right)
\Bigg]^{-1}.
\end{aligned}
\end{equation}

For weak but finite inertia, the wall correction must additionally
account for the inertia-dependent modification of the disturbance flow
\cite{Tashibana.1976,Noichl.2026}. Based on model (2), the wall-bounded
velocity in the small-angle regime is approximated as
\begin{equation}
\label{eq16}
\begin{aligned}
\left.
U_{\mathrm{th,wb}}^{Re\leq1}(\phi)
\right|_{\mid\shortmid}^{(2)}
\approx{}&
\frac{
\left(\rho_{\mathrm{P}}-\rho\right)r^2g
}{
2\eta
F_{\mathrm{D,ub}}^{\ast,\theta=0^\circ,\,
2Re_{\mathrm{P}}=O(1)}
\sec\phi
}
\\
&\times
\Bigg[
1
+\frac{3}{8}Re_{\mathrm{P,th}}(\phi)
\\
&\qquad
+\frac{9}{16}
F_{\mathrm{D,ub}}^{\ast,\theta=0^\circ,\,
2Re_{\mathrm{P}}=O(1)}
\sec\phi
\left(\frac{l}{d}\right)
\mathcal{K}(\phi)
\Bigg]^{-1}.
\end{aligned}
\end{equation}

The corresponding theoretical particle Reynolds number is
\begin{equation}
\label{eq17}
Re_{\mathrm{P,th}}(\phi)
=
\frac{
\left(\rho_{\mathrm{P}}-\rho\right)\rho lr^2g\cos\phi
}{
2\eta^2
F_{\mathrm{D,ub}}^{\ast,\theta=0^\circ,\,
2Re_{\mathrm{P}}=O(1)}
}.
\end{equation}
The inertia-dependent wall function is defined as
\begin{equation}
\label{eq18}
\mathcal{K}(\phi)
=
1
-\frac{4}{3}
\left[
\frac{
Re_{\mathrm{P,th}}(\phi)
}{
2(l/d)
}
\right]
+\frac{23}{16}
\left[
\frac{
Re_{\mathrm{P,th}}(\phi)
}{
2(l/d)
}
\right]^3.
\end{equation}

As shown below, Brenner's model (1) provides a more accurate description
of the cylinder velocity at larger wall distances, \(d/l>2\). Including
the third-order wall correction yields
\begin{equation}
\label{eq19}
\begin{aligned}
\left.
U_{\mathrm{th,wb}}^{Re\leq1}(\phi)
\right|_{\mid\shortmid}^{(1)}
\approx{}&
\frac{
\left(\rho_{\mathrm{P}}-\rho\right)r^2g
}{
2\eta
F_{\mathrm{D,ub}}^{\ast,\theta=0^\circ,\,
2Re_{\mathrm{P}}=O(1)}
\sec\phi
}
\\
&\times
\Bigg[
1
-\frac{3}{8}Re_{\mathrm{P,th}}(\phi)
\\
&\qquad
-\frac{9}{16}
F_{\mathrm{D,ub}}^{\ast,\theta=0^\circ,\,
2Re_{\mathrm{P}}=O(1)}
\sec\phi
\left(\frac{l}{d}\right)
\mathcal{K}(\phi)
\\
&\qquad
+\frac{1}{8}
F_{\mathrm{D,ub}}^{\ast,\theta=0^\circ,\,
2Re_{\mathrm{P}}=O(1)}
\sec\phi
\left(\frac{l}{d}\right)^3
\Bigg].
\end{aligned}
\end{equation}

Equations~\eqref{eq16} and \eqref{eq19} provide approximations for the
wall-bounded translational velocity at weak but finite inertia in the
regime of small to moderate orientation and trajectory angles. They are
based on the small-angle approximation introduced in Eq.~\eqref{eq9},
while additionally accounting for the inertia-dependent modification of
the wall correction.

To extend the velocity model to the broader range of orientations
encountered experimentally, the two limiting cylinder configurations are
considered separately. For the wall-parallel configuration,
\(\theta=0^\circ\), the wall-bounded velocity
\(U_{\mathrm{th,wb}}^{\theta=0^\circ,\,Re\lesssim1}(\chi,d/l)\)
is calculated using the inertia-corrected axial drag coefficient given by
Eq.~\eqref{eq5}. For the transverse configuration,
\(\theta=90^\circ\), the corresponding velocity
\(U_{\mathrm{th,wb}}^{\theta=90^\circ,\,Re\lesssim1}(\chi,d/l)\)
is calculated analogously using the inertia-corrected transverse drag
coefficient given by Eq.~\eqref{eq6}. In both limiting configurations,
the corresponding inertia-corrected wall factor is applied.

The velocity at an arbitrary orientation is subsequently approximated by
projecting these two limiting wall-bounded velocities onto the
instantaneous cylinder orientation. This construction constitutes a
kinematic interpolation between the two limiting resistance
configurations rather than an exact wall-bounded resistance relation for
an arbitrarily oriented cylinder. It preserves the respective limiting
velocities at \(\theta=0^\circ\) and \(\theta=90^\circ\), while providing
a continuous approximation for intermediate orientations.

The contribution associated with the wall-parallel limiting
configuration is approximated as
\begin{equation}
\label{eq20}
U_{\parallel,\mathrm{th,wb}}(\theta)
=
U_{\mathrm{th,wb}}^{\theta=0^\circ,\,Re\lesssim1}
\left(\chi,\frac{d}{l}\right)
\cos\theta,
\end{equation}
whereas the contribution associated with the transverse limiting
configuration is
\begin{equation}
\label{eq21}
U_{\perp,\mathrm{th,wb}}(\theta)
=
U_{\mathrm{th,wb}}^{\theta=90^\circ,\,Re\lesssim1}
\left(\chi,\frac{d}{l}\right)
\sin\theta.
\end{equation}

The magnitude of the theoretical wall-bounded translational velocity is
then obtained from the vector sum of both contributions,
\begin{equation}
\label{eq22}
\begin{aligned}
U_{\mathrm{th,wb}}(\theta)
={}&
\Bigg\{
\left[
U_{\mathrm{th,wb}}^{\theta=0^\circ,\,Re\lesssim1}
\left(\chi,\frac{d}{l}\right)
\cos\theta
\right]^2
\\
&\qquad+
\left[
U_{\mathrm{th,wb}}^{\theta=90^\circ,\,Re\lesssim1}
\left(\chi,\frac{d}{l}\right)
\sin\theta
\right]^2
\Bigg\}^{1/2}.
\end{aligned}
\end{equation}

A direct insertion of the projected transverse force,
\(F_{\perp}(\chi,\theta)
=F_{\perp}^{\theta=90^\circ}\sin\theta\),
into the wall-bounded force balance is not used. Solving such a force
balance for the velocity introduces a factor proportional to
\(1/\sin\theta\), which becomes singular as
\(\theta\rightarrow0^\circ\), although the physical translational
velocity remains finite. The interpolation defined by
Eqs.~\eqref{eq20}--\eqref{eq22} avoids this artificial singularity and
continuously recovers the corresponding limiting velocities for
\(\theta\rightarrow0^\circ\) and \(\theta\rightarrow90^\circ\).

\subsection{Angular velocity}
\label{sec2_2}

The theoretical angular velocity of a cylinder rotating within the
\(x\)-\(y\) plane follows from the torque balance about an axis through
the center of mass perpendicular to the plane of motion, \textit{i.e.}
the \(z\)-axis in Fig.~\ref{Fig1}. The angular velocity is defined as
\(\omega=\mathrm{d}\theta/\mathrm{d}t\). The rotational equation of
motion reads
\begin{equation}
\label{eq23}
I_{\mathrm{P}}\dot{\omega}
=
T_{\mathrm{H}}
+
T_{\mathrm{I}}
+
T_{\mathrm{W}},
\end{equation}
where \(I_{\mathrm{P}}\) is the moment of inertia of the cylinder,
\(T_{\mathrm{H}}\) is the hydrodynamic resistance torque associated with
the rotation of the cylinder, \(T_{\mathrm{I}}\) denotes the
inertia-induced torque resulting from its translational motion in an
unbounded fluid, and \(T_{\mathrm{W}}\) represents the additional torque
arising from the hydrodynamic interaction with the wall.

Analogous to the hydrodynamic drag in the translational force balance,
the hydrodynamic resistance torque opposes the rotational motion and can
be written as
\begin{equation}
\label{eq24}
T_{\mathrm{H}}
=
-C_{\mathrm{H}}\omega,
\end{equation}
where \(C_{\mathrm{H}}\) denotes the rotational resistance coefficient.
Under the quasi-steady assumption, rotational particle inertia is
neglected, \(I_{\mathrm{P}}\dot{\omega}\approx0\), such that the angular
velocity follows from
\begin{equation}
\label{eq25}
\omega
=
\frac{
T_{\mathrm{I}}+T_{\mathrm{W}}
}{
C_{\mathrm{H}}
}.
\end{equation}

\subsubsection{Rotational resistance}
\label{sec2_2_1}

For a slender cylinder rotating about an axis perpendicular to its
symmetry axis in an unbounded fluid, the leading-order rotational
resistance follows from slender-body theory. Using the present
definition of the cylinder aspect ratio, $\chi=l/r$, the corresponding
slenderness parameter is
\begin{equation}
\label{eq26}
\epsilon
=
\frac{1}{\ln(2\chi)}.
\end{equation}
The leading-order rotational resistance coefficient can then be written
as~\cite{Batchelor.1970}
\begin{equation}
\label{eq27}
C_{\mathrm{H}}^{(1)}
=
\frac{8\pi}{3}
\eta l^3\epsilon.
\end{equation}
Equation~\eqref{eq27} represents the leading-order slender-body
resistance and is recovered in the limit of large aspect ratio.

For cylinders of finite aspect ratio, higher-order contributions to the
rotational resistance become relevant. Pierson \textit{et al.}
\cite{Pierson.2021} extended the slender-body expansion of the
rotation-induced hydrodynamic torque to fourth order in the small
parameter $\epsilon$. Their fourth-order expression includes an
empirical aspect-ratio-dependent prefactor to improve the approximation
for cylinders with moderate aspect ratios. 

Using the present definition of the total cylinder length $2l$, the
rotational resistance is written as
\begin{equation}
\label{eq28}
C_{\mathrm{H}}
=
\frac{8\pi}{3}
\eta l^3
\mathcal{C}(\chi),
\end{equation}
where the fourth-order approximation reads
\begin{equation}
\label{eq29}
\begin{aligned}
\mathcal{C}(\chi)
={}&
\epsilon
+
\epsilon^2
\left(
\frac{11}{6}-\ln 2
\right)
\\
&+
\epsilon^3
\left[
\frac{161}{36}
-\frac{\pi^2}{12}
-\frac{11}{3}\ln 2
+(\ln 2)^2
\right]
\\
&+
\epsilon^4
\left(1-\frac{1}{(2\chi)^{1.2}}\right)^5
\left[
-\frac{5}{4}\zeta(3)
+\frac{1033}{72}
-\ln^3 2
+\frac{11}{2}\ln^2 2
\right.
\\
&\qquad\left.
-\frac{161}{12}\ln 2
-\pi^2
\left(
\frac{11}{24}
-\frac{1}{8}\ln 4
\right)
\right].
\end{aligned}
\end{equation}
The first term in Eq.~\eqref{eq29} recovers the leading-order result in
Eq.~\eqref{eq27}, whereas the additional terms account for higher-order
contributions to the rotational resistance associated with a finite
aspect ratio.

\subsubsection{Inertia-induced rotation in an unbounded fluid}
\label{sec2_2_2}

In addition to the rotational resistance torque, translational fluid
inertia generates an orienting torque on an inclined cylinder even in
an unbounded fluid. Khayat and Cox~\cite{Khayat.1989} derived this
contribution for slender rods at low Reynolds numbers. Higher-order
expressions applicable to cylinders of moderate aspect ratio and finite
Reynolds number were subsequently developed by Kharrouba \textit{et al.}
\cite{Kharrouba.2021}.

For the present definitions of the orientation angle \(\theta\) and
trajectory angle \(\phi\), the instantaneous incidence angle between
the cylinder axis and its translational velocity is defined as
\begin{equation}
\label{eq30}
\alpha
=
\theta-\phi.
\end{equation}

Using the correlation of Kharrouba \textit{et al.}
\cite{Kharrouba.2021}, and expressing their total cylinder length as
\(2l\), the inertia-induced torque becomes
\begin{equation}
\label{eq31}
\begin{aligned}
T_{\mathrm{I,K}}
={}&
-\frac{5\pi}{12}
\eta U l^2
\frac{Re}
{\sqrt{1+\chi Re^{1.1}}}
\\
&\times
\left[
\frac{\chi}{(\ln\chi)^2}
+
\frac{
13.5-30Re^{1/2}
}{
\chi(\ln\chi)^3
}
e^{-0.7Re}
\right]
\sin(2\alpha),
\end{aligned}
\end{equation}
where \(U=\sqrt{U_{\mathrm{P,x}}^2+U_{\mathrm{P,y}}^2}\) is the
instantaneous translational velocity magnitude and \(Re\) is the
diameter-based Reynolds number defined in Sec.~\ref{sec2_1}.

Combining the inertia-induced torque in Eq.~\eqref{eq31} with the
third-order rotational resistance in Eq.~\eqref{eq28} yields the
theoretical angular velocity of the translating cylinder in an
unbounded fluid,
\begin{equation}
\label{eq32}
\omega_{\mathrm{th,ub}}^{\mathrm{KP}}
=
\frac{
T_{\mathrm{I,K}}
}{
C_{\mathrm{H}}
}.
\end{equation}
The superscript \(\mathrm{KP}\) denotes the combination of the
inertia-induced torque correlation of Kharrouba \textit{et al.}
\cite{Kharrouba.2021} with the rotational resistance of Pierson
\textit{et al.}~\cite{Pierson.2021}.

\subsubsection{Wall-induced rotation}
\label{sec2_2_3}

The presence of a plane wall introduces an additional
translation--rotation coupling. Russel \textit{et al.}
\cite{RusselWB.1977} investigated this interaction for a slender
cylinder sedimenting close to a vertical plane wall using slender-body
theory together with the corresponding hydrodynamic image system. Their
resulting analytical expression for the angular velocity depends on
both the cylinder--wall distance and the cylinder orientation.

In the present model, only the limiting solution for a cylinder oriented
parallel to the wall, \(\theta=0^\circ\), is retained to describe the
wall-induced contribution. The general expression presented by Russel
\textit{et al.}~\cite{RusselWB.1977} in their Eq.~(8) contains a term
denoted by \(F\), whose relation to the quantities \(F_{i}\)
appearing elsewhere in their derivation is not unambiguous.
Although this may indicate a typographical inconsistency in the
published expression, this interpretation cannot be verified
conclusively. The published general expression is therefore not
modified here.

Furthermore, the cylinders considered in the present study have
moderate aspect ratios and are consequently far from the asymptotically
slender limit underlying the analytical treatment of Russel
\textit{et al.}~\cite{RusselWB.1977}. For this reason, the
\(\theta=0^\circ\) solution is used here as the leading-order
representation of the wall-distance dependence of the
translation--rotation coupling. This choice should not be interpreted
as implying that the wall-induced torque is generally independent of
the cylinder orientation. Rather, the additional orientation-dependent
terms of the complete slender-body expression are not retained in the
present model.

In the notation of Russel \textit{et al.}~\cite{RusselWB.1977}, the
dimensionless center--wall distance is \(X=d/l\). For
\(\theta=0^\circ\), the corresponding wall function is
\begin{equation}
\label{eq33}
B(X,0)
=
4X
\left(
\sqrt{X^2+1}
-
X
\right)
-
\frac{2X}{\sqrt{X^2+1}},
\qquad
X=\frac{d}{l}.
\end{equation}
Hereafter, the wall distance is expressed directly as \(d/l\),
consistently with Sec.~\ref{sec2_1}.

Using the sign convention for \(\theta\) adopted in the present study,
the leading-order wall-induced angular velocity obtained from Russel
\textit{et al.} becomes
\begin{equation}
\label{eq34}
\omega_{\mathrm{wall,R}}
=
-\frac{
3F_0
}{
32\pi\eta l^2
}
B\left(
\frac{d}{l},0
\right),
\end{equation}
where
\begin{equation}
\label{eq35}
F_0
=
\left(
\rho_{\mathrm{P}}-\rho
\right)
V_{\mathrm{P}}g
\end{equation}
is the net gravitational force acting on the cylinder. The subscript
\(\mathrm{wall,R}\) denotes the wall-induced contribution obtained from
the theory of Russel \textit{et al.}~\cite{RusselWB.1977}.

Equation~\eqref{eq34} is based on a leading-order slender-body
approximation. Russel \textit{et al.}~\cite{RusselWB.1977} themselves
noted that higher-order corrections to the resistance coefficients can
constitute an important source of error in calculated trajectories.
This becomes particularly relevant for the present cylinders, whose
aspect ratios are considerably smaller than those associated with the
asymptotically slender limit.

The relation between the leading-order Russel wall contribution and the
finite-aspect-ratio rotational resistance can be made explicit by
rewriting Eq.~\eqref{eq34} in terms of a torque. Multiplication by the
leading-order rotational resistance coefficient in Eq.~\eqref{eq27}
gives
\begin{equation}
\label{eq36}
\begin{aligned}
T_{\mathrm{W,R}}^{(1)}
&=
C_{\mathrm{H}}^{(1)}
\omega_{\mathrm{wall,R}}
\\
&=
-\frac{
F_0l
}{4}
\epsilon
B\left(
\frac{d}{l},0
\right).
\end{aligned}
\end{equation}
Equation~\eqref{eq36} thus represents the leading-order wall-induced
torque underlying the angular-velocity expression of Russel
\textit{et al.}~\cite{RusselWB.1977}.

To account for the finite aspect ratio of the cylinders investigated
here, the leading-order wall-coupling term in Eq.~\eqref{eq36} is
retained, while the leading-order rotational resistance
\(C_{\mathrm{H}}^{(1)}\) is replaced by the third-order rotational
resistance \(C_{\mathrm{H}}\) of Pierson \textit{et al.}
\cite{Pierson.2021}. The resulting wall-induced angular velocity is
\begin{equation}
\label{eq37}
\omega_{\mathrm{th,wb}}^{Re\approx0}
=
\frac{
T_{\mathrm{W,R}}^{(1)}
}{
C_{\mathrm{H}}
}
=
\omega_{\mathrm{wall,R}}
\frac{
\epsilon
}{
\mathcal{C}(\chi)
}.
\end{equation}

Equation~\eqref{eq37} is therefore interpreted as a
finite-aspect-ratio-corrected leading-order wall contribution. It does
not constitute a complete finite-aspect-ratio solution of the
wall-bounded hydrodynamic problem. Instead, it assumes that the dominant
finite-\(\chi\) correction can be represented through the rotational
resistance of the cylinder, while the leading-order wall-coupling
function of Russel \textit{et al.}~\cite{RusselWB.1977} is retained.
Higher-order wall corrections and a possible wall-induced modification
of the rotational resistance itself are consequently not included.

\subsubsection{Combined angular-velocity model}
\label{sec2_2_4}

At weak but finite inertia, the theoretical angular velocity is obtained
by combining the inertia-induced torque in the unbounded fluid with the
wall-induced torque. The quasi-steady torque balance becomes
\begin{equation}
\label{eq38}
T_{\mathrm{I,K}}
+
T_{\mathrm{W,R}}^{(1)}
-
C_{\mathrm{H}}\omega
=
0.
\end{equation}

Accordingly,
\begin{equation}
\label{eq39}
\omega_{\mathrm{th,wb}}^{Re\lesssim1}
=
\frac{
T_{\mathrm{I,K}}
+
T_{\mathrm{W,R}}^{(1)}
}{
C_{\mathrm{H}}
}.
\end{equation}

Using Eqs.~\eqref{eq32} and \eqref{eq37}, the combined model can
equivalently be written in the form used for comparison with the
experimental results,
\begin{equation}
\label{eq40}
\omega_{\mathrm{th,wb}}^{Re\lesssim1}
=
\omega_{\mathrm{th,wb}}^{Re\approx0}
+
\omega_{\mathrm{th,ub}}^{\mathrm{KP}}.
\end{equation}

The first contribution,
\(\omega_{\mathrm{th,wb}}^{Re\approx0}\), represents the wall-induced
rotation in the creeping-flow limit, including the finite-aspect-ratio
correction to the rotational resistance according to Eq.~\eqref{eq37}.
The second contribution,
\(\omega_{\mathrm{th,ub}}^{\mathrm{KP}}\), represents the additional
inertia-induced rotation predicted for a translating cylinder in an
unbounded fluid according to Eq.~\eqref{eq32}. No empirical fitting
parameter is introduced in either contribution.

\section{Methods}
\label{sec3}

\subsection{Experimental setup and definitions}
\label{sec3_1}

\begin{figure}
    \centering
    \includegraphics[width=0.8\textwidth]{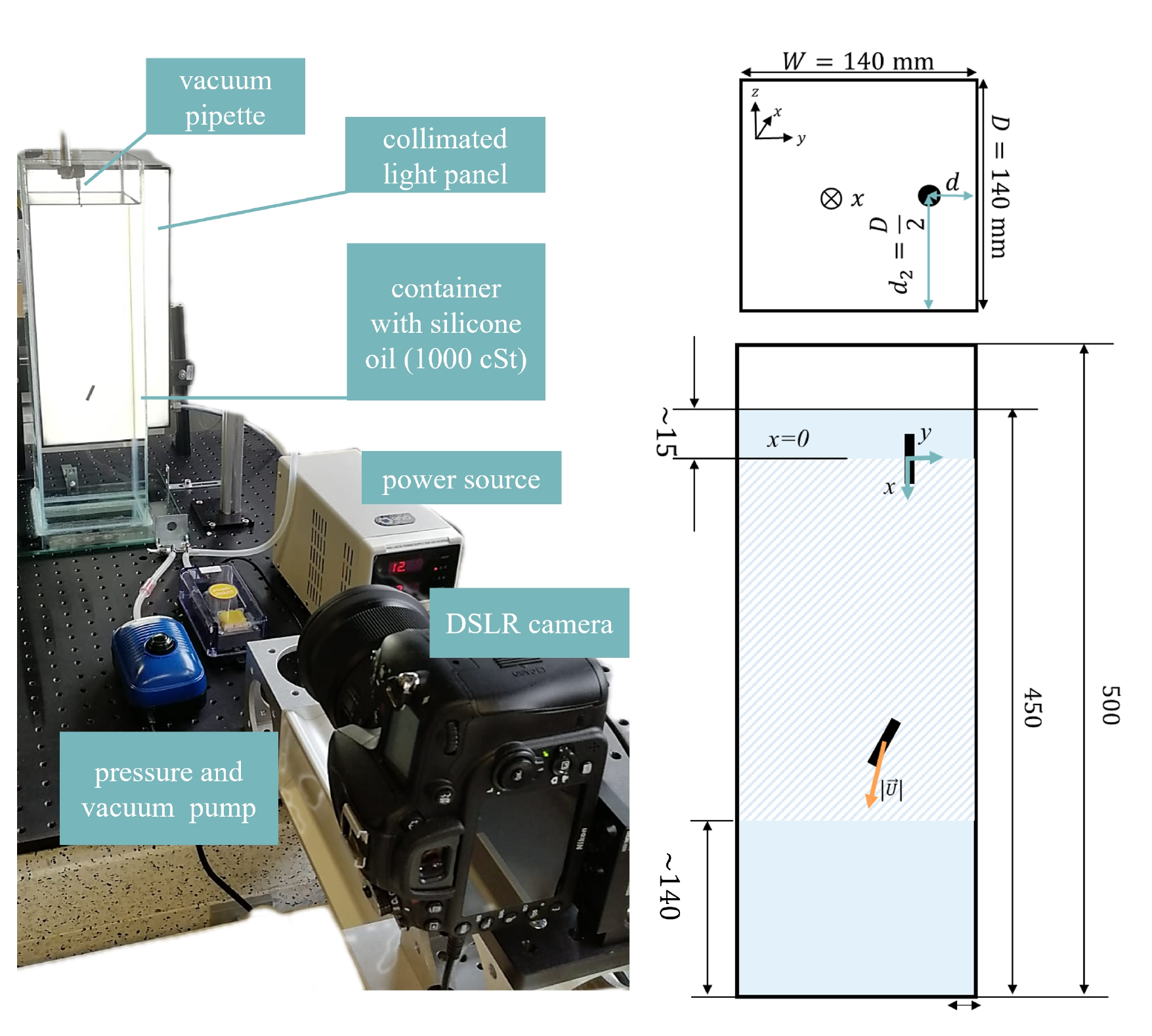}
    \caption{Experimental setup and measurement geometry. Left: photograph
    of the experimental setup, comprising a glass container filled with
    silicone oil of nominal kinematic viscosity \(1000~\mathrm{cSt}\), a
    collimated LED light panel with its power supply, a vacuum pipette
    connected to a vacuum/pressure pump, and a DSLR camera. Upper right: top
    view of the square container cross-section with width
    \(W=140~\mathrm{mm}\) and depth \(D=140~\mathrm{mm}\). The cylinder is
    positioned in the mid-plane between the front and rear walls,
    \(d_2=D/2\), while \(d\) denotes the distance from the nearest lateral
    wall. Lower right: side view corresponding to the imaging
    (\(xy\))-plane. The container height is \(500~\mathrm{mm}\), and the
    liquid filling height is approximately \(450~\mathrm{mm}\). The vertical
    coordinate is referenced to the initial particle-center position,
    \(x=0\), located approximately \(15~\mathrm{mm}\) below the free surface.
    The hatched region indicates the evaluated measurement region, extending
    from \(x=0\) to approximately \(140~\mathrm{mm}\) above the container
    bottom.}
    \label{Fig2}
\end{figure}

Sedimentation experiments with cylindrical model particles were performed in
the same rectangular duct previously used for experiments with spherical
model particles~\cite{Noichl.2022,Noichl.2026}. Figure~\ref{Fig2} shows a
photograph of the experimental setup together with top- and side-view
schematics illustrating the container geometry, the particle position, the
coordinate system, and the evaluated measurement region.

A glass container with internal dimensions of
\(140~\mathrm{mm}\times140~\mathrm{mm}\times500~\mathrm{mm}\)
(\(W\times D\times H\)) was used for all experiments. The container was
filled with silicone oil to a height of approximately \(450~\mathrm{mm}\).
The silicone oil had a nominal kinematic viscosity of
\(1000~\mathrm{cSt}\). Its dynamic viscosity was measured using a rotational
viscometer (IKA\textsuperscript{\textregistered} Rotavisc lo-vi), yielding
\(\eta=0.9797\pm0.012~\mathrm{Pa\,s}\). The liquid density at room
temperature was determined using a hydrostatic weighing system consisting of
a Sartorius\textsuperscript{\textregistered} ENTRIS BCE 224i-1s analytical
balance equipped with a Sartorius\textsuperscript{\textregistered} YDK03
density determination kit. The measured density was
\(\rho=971.28\pm0.134~\mathrm{kg\,m^{-3}}\).

The selected liquid properties enabled sedimentation experiments in the
creeping-flow regime, \(Re_{\mathrm{P}}\lesssim0.1\). Consequently,
centimetre-sized model particles with only a small density difference
relative to the surrounding fluid experienced flow conditions comparable to
those of micrometre-sized particles sedimenting in aqueous liquids.

Before release, each rod was held by a pipette under weak vacuum and immersed
in the liquid. The orientation of the pipette allowed the cylinder to be
positioned at different preset angles prior to release. However, this preset
angle did not correspond directly to the experimentally determined initial
orientation angle \(\theta_{\mathrm{ini}}\). Upon release from rest, the
cylinders exhibited a pronounced initial reorientation. As liquid penetrated
the gap between the cylinder and the pipette tip during detachment, the
cylinder experienced a torque towards smaller orientation angles.
Consequently, the first measurable orientation angle
\(\theta_{\mathrm{ini}}\), defined at the instant when the cylinder had
completely detached from the pipette, was substantially smaller than the
preset angle. For example, the maximum experimentally observed initial
orientation of \(\theta_{\mathrm{ini}}\approx49^\circ\) required a preset
angle of approximately \(65^\circ\). Larger preset angles could not be
achieved with the present release mechanism because the pipette holder could
not be positioned at a more horizontal orientation.

As illustrated in the top view of Fig.~\ref{Fig2}, the rod was positioned in
the mid-plane between the front and rear walls, \(d_2=D/2\), while the
distance \(d\) from the nearest lateral wall was varied. After positioning,
the vacuum was released and the cylindrical model particle was allowed to
sediment under gravity. The cylinder was thus released with negligible
initial translational velocity, \(U_{\mathrm{P}}(t=0)\approx0\), into an
initially quiescent fluid,
\(\boldsymbol{u}_{\mathrm{f}}(t=0)
=(0,0,0)^{\mathrm{T}}\).

The instant of complete detachment from the pipette defines the beginning of
the evaluated particle trajectory. The particle-center position at this
instant is used as the origin of the vertical coordinate, \(x=0\), as
indicated by the horizontal line located approximately \(15~\mathrm{mm}\)
below the free surface in the side view of Fig.~\ref{Fig2}.

The sedimentation experiments were recorded using a DSLR camera
(Nikon\textsuperscript{\textregistered} D7200 equipped with a
Sigma\textsuperscript{\textregistered} 50~mm \(f/1.4\) lens). Videos were
recorded at a resolution of \(1920\times1080~\mathrm{px}\) and a frame rate
of \(25~\mathrm{fps}\).

During the experiments, the container was illuminated from the side opposite
the camera using a collimated LED light panel. This backlighting arrangement
produced a sharp particle silhouette on the camera sensor. To minimise image
distortion, for example due to lens aberrations, the camera was aligned such
that the particle trajectory remained close to the centre of the recorded
image.

Before each experiment, a high-resolution reference image with a resolution
of \(3368\times6000~\mathrm{px}\) was recorded. The cylindrical tube of the
vacuum pipette, whose outer diameter was known, served as a reference length
for determining the pixel-to-millimetre conversion. Using this calibration,
the actual length and diameter of the individual cylinder used in each
experiment were determined from the corresponding dimensions measured in the
reference image. The spatial and angular quantities used for the evaluation
are defined in Fig.~\ref{Fig1} and Sec.~\ref{sec2}.

The focal plane of the camera coincided with the \(xy\)-plane of the
experimental setup. When the cylinder axis remained within the focal plane
and the cylinder rotated about the \(z\)-axis through the orientation angle
\(\theta\), the centre of mass coincided with the centroid of the projected
particle area \(A_{\mathrm{px}}\). For a homogeneous cylinder, the centroid
is located at the axial half-length \(l\) and at the radial midpoint \(r\).

Only cylinders rotating within the imaging plane were considered in the
present study. Consequently, only rotation about the \(z\)-axis was
evaluated. The particle trajectories, orientation angles, and translational
velocities were determined using a custom image-processing tool developed in
MATLAB\textsuperscript{\textregistered}.

\subsection{Model-particle fabrication}
\label{sec3_2}

The cylindrical model particles were fabricated from epoxy resin.
Weicon\textsuperscript{\textregistered} MS 1000 casting resin was prepared
according to the manufacturer's specifications and coloured black by adding
\(1~\mathrm{wt\text{-}\%}\) iron oxide powder.

The liquid epoxy resin mixture was drawn into \(1~\mathrm{mL}\) syringes
(Ominfix\textsuperscript{\textregistered}-F LuerSolo,
\(0.01\)--\(1~\mathrm{mL}\)). Each syringe barrel was completely filled with
the resin mixture before being placed vertically with the needle adapter
pointing upwards. The vertical orientation allowed air trapped in the
barrel, as well as air bubbles remaining from the mixing process, to rise and
escape through the needle adapter. The syringes were fixed in a suitable
holder and cured in an oven according to the manufacturer's specifications.

After curing, the needle adapter was removed using a tube cutter, thereby
opening the syringe barrel completely. The plunger was removed, and the space
between the cured polymer and the syringe flange was filled with distilled
water. Subsequently, the plunger was reinserted and advanced, forcing water
between the polymer cylinder and the syringe wall. This released the cured
cylinder from the syringe wall and allowed it to be pressed out through the
opened barrel.

The resulting cylinders had a diameter of approximately
\(2r\approx4.7~\mathrm{mm}\). The long cylinders were cut into shorter rods
with total lengths of approximately \(2l\approx18~\mathrm{mm}\),
\(2l\approx24~\mathrm{mm}\), and \(2l\approx28~\mathrm{mm}\),
corresponding to aspect ratios of \(\bar{\chi}\approx3.8\), \(\bar{\chi}\approx5.2\),
and \(\bar{\chi}\approx5.8\), respectively. After cutting, both end faces of each
rod were carefully flattened using fine-grit sandpaper to obtain flat
surfaces perpendicular to the cylinder axis. This finishing step ensured
well-defined cylinder lengths and planar end faces, thereby minimising
geometric deviations between individual model particles.

The cured epoxy resin had a measured solid density of
\(\rho_{\mathrm{s,Epoxy}}=1159.8\pm1.9~\mathrm{kg\,m^{-3}}\) and behaved as
a rigid material. According to the manufacturer's specifications, the cured
epoxy resin had a Young's modulus of \(E=2.9~\mathrm{GPa}\). The solid
density was determined by hydrostatic weighing using a
Sartorius\textsuperscript{\textregistered} ENTRIS BCE 224i-1s analytical
balance equipped with a Sartorius\textsuperscript{\textregistered} YDK03
density determination kit. The balance had a readability of
\(0.1~\mathrm{mg}\), and all density measurements were performed in
distilled water.

\section{Results and Discussion}
\label{sec4}

The results are presented in three steps. First, the sedimentation of
axially aligned cylinders in the center of the rectangular duct is examined
to establish the reference behavior in the absence of a dominant interaction
with a single lateral wall. The reproducibility of the sedimentation
kinematics and the influence of the cylinder aspect ratio are assessed, and
the measured steady velocities are compared with the theoretical predictions
introduced in Sec.~\ref{sec2_1}. Subsequently, the translational kinematics
of initially inclined cylinders sedimenting at different distances from a
plane wall are investigated. Particular attention is paid to the coupled
evolution of the cylinder orientation, wall distance, and translational
velocity and to the applicability of the wall-corrected velocity models.
Finally, the corresponding angular kinematics are analyzed and compared with
the wall-induced and inertia-corrected angular-velocity models developed in
Sec.~\ref{sec2_2}. Additional results for cylinders with the smaller aspect
ratio \(\chi\approx3.8\) are provided in Appendix~\ref{secA1}.

Throughout this section, individual experimental runs are shown rather than
data averaged over repeated measurements. This distinction is important
because the sedimentation dynamics considered here are sensitive to the
instantaneous particle configuration. 
Small and experimentally unavoidable variations in the initial conditions,
including the initial orientation \(\theta_{\mathrm{ini}}\), the initial wall
distance \(d_{\mathrm{ini}}/l\), and small differences in particle properties
between individual specimens, can lead to different subsequent evolutions of
the particle orientation, wall distance, and translational and angular
velocities.
These differences therefore do not necessarily represent random experimental scatter that can be removed by ensemble averaging. In particular, averaging absolute
velocities or combining trajectories obtained at slightly different initial
conditions would destroy the direct correspondence between
\(\theta\), \(d/l\), \(U_{\mathrm{P}}\), and \(\omega\) along an individual
trajectory and may consequently obscure systematic features of the
particle dynamics. Repeated experiments are instead used to assess the
reproducibility of the observed behavior, while the individual trajectories
and their associated kinematic quantities are retained for the analysis.
Numerical scatter arising from the evaluation of the recorded trajectories,
particularly from temporal differentiation, was minimized by defining the
evaluation time step prior to the analysis and applying it consistently to
the experimental data.

\subsection{Axially aligned sedimentation of short cylinders in the center
of a rectangular duct}
\label{sec4_1}
The sedimentation behavior in the center of the rectangular duct was first
examined for axially aligned cylinders with mean aspect ratios of
\(\bar{\chi}\approx5.2\) and \(\bar{\chi}\approx5.8\).
This configuration represents the symmetric case, in which the four
surrounding container walls are approximately equidistant from the particle
and therefore exert comparable, yet still measurable, influences on its
sedimentation kinematics. It provides a reference for the subsequent
experiments in which the symmetry is broken by positioning the cylinder
closer to one of the lateral walls.

\begin{figure}
    \centering
    \includegraphics[width=0.7\textwidth]{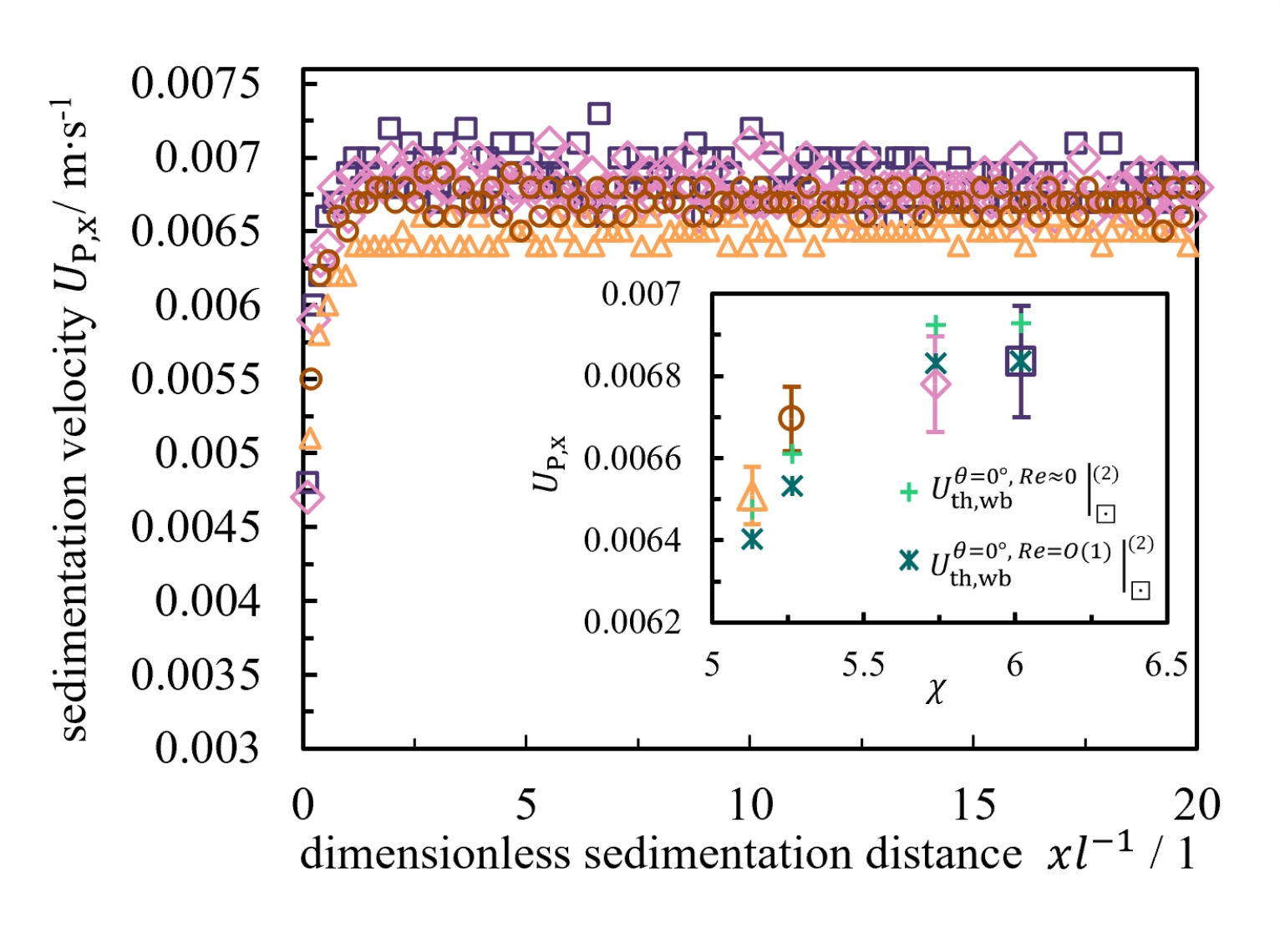}
    \caption{Sedimentation velocity \(U_{\mathrm{P,x}}\) plotted against the dimensionless sedimentation distance \(x/l\) for cylinders (\(E\approx2900\)~MPa; \(\rho_{\mathrm P}\approx1159.8\)~kg\,m\(^{-3}\)) belonging to two groups with mean aspect ratios \(\bar{\chi}\approx5.2\) and \(\bar{\chi}\approx5.8\), released with negligible initial translational velocity in an approximately axially aligned orientation (\(\theta_{\mathrm{ini}}\approx0^\circ\)) at \(Re_{\mathrm P}\approx0.1\) in the center of the duct. \textbf{Inset:} Mean steady sedimentation velocity and corresponding standard deviation for each experiment plotted against the individually measured aspect ratio \(\chi\). Colors and symbols correspond to those used for the respective experiments in the main panel. The experimental results are compared with the theoretical sedimentation velocities calculated using Ui's model (2), Eq.~\eqref{eq13}, with the creeping-flow axial drag of Eq.~\eqref{eq3} (light-green plus symbols) and with the inertia-corrected axial drag of Eq.~\eqref{eq5} (dark-green cross symbols).}
    \label{Fig3}
\end{figure}

Figure~\ref{Fig3} shows the sedimentation velocity
\(U_{\mathrm{P,x}}\) as a function of the dimensionless sedimentation distance
\(x/l\) for cylinders with \(\bar{\chi}\approx5.2\) and
\(\bar{\chi}\approx5.8\), released with negligible initial translational
velocity in an approximately axially aligned orientation,
\(\theta_{\mathrm{ini}}\approx0^\circ\), in the center of the rectangular
duct.

The velocity curves in Fig.~\ref{Fig3} show that the axially aligned cylinders
initially accelerate before approaching an approximately steady sedimentation
velocity. The corresponding spatial and angular trajectories (not shown)
exhibit similarly stable behavior when the cylinders are released close to
the center of the duct with \(\theta_{\mathrm{ini}}\approx0^\circ\): both the
orientation angle \(\theta\) and the dimensionless wall distance \(d/l\)
remain approximately constant throughout the evaluated sedimentation path.
Small deviations from this symmetric initial configuration, either due to a
non-zero initial orientation or a slightly off-center release position,
result in gradual changes in the orientation angle. Depending on the
orientation of the cylinder relative to the nearest wall, \(\theta\) may
either increase or decrease during sedimentation.

In addition, a focusing behavior is consistently observed across the
experiments. Cylinders released slightly off-center tend to migrate toward an
equilibrium position within the duct as sedimentation proceeds. This tendency
is accompanied by an apparent dependence on cylinder size: the longer
cylinders investigated here show a stronger tendency to recover an axially
aligned orientation and a centered position, whereas the shorter cylinders
exhibit larger deviations in orientation and lateral position in response to
small differences in their initial conditions.

The inset of Fig.~\ref{Fig3} shows the mean steady sedimentation velocity
obtained from each experiment in the main panel plotted against the
individually measured aspect ratio \(\chi\), which were \(\chi\approx 5.1;5.3;5.7;6.0\). The experimental values are
compared with the theoretical velocities calculated using Ui's model (2),
Eq.~\eqref{eq13}, with either the creeping-flow axial drag of
Eq.~\eqref{eq3} or the inertia-corrected axial drag of Eq.~\eqref{eq5}.
Overall, the measured velocities are in good agreement with the predictions
based on Ui's model. For the cylinders with the larger aspect ratio, the
experimental values are more closely described when the inertia correction
is included, whereas the difference between the two theoretical predictions
is less pronounced for the cylinders with the smaller aspect ratio. This
trend is consistent with the slightly larger particle Reynolds numbers of
the longer cylinders, with the mean \(Re_{\mathrm{P}}\) increasing from
approximately \(0.08\) for the \(\bar{\chi}\approx5.2\) group to
approximately \(0.09\) for the \(\bar{\chi}\approx5.8\) group.

For comparison, the theoretical velocities were also evaluated using
Brenner's model (1), Eq.~\eqref{eq12}, which consistently yielded lower
velocities than those measured experimentally and is therefore not included
in the inset. The origin of this systematic underprediction cannot be
unambiguously identified from the present experiments. Differences in the
higher-order treatment of the wall correction may contribute to the
discrepancy; however, a quantitative assessment of the individual
contributions is beyond the scope of the present study.

\subsection{Near-wall translational kinematics of initially inclined cylinders}
\label{sec4_2}

\begin{figure}
    \centering
    \includegraphics[width=0.65\textwidth]{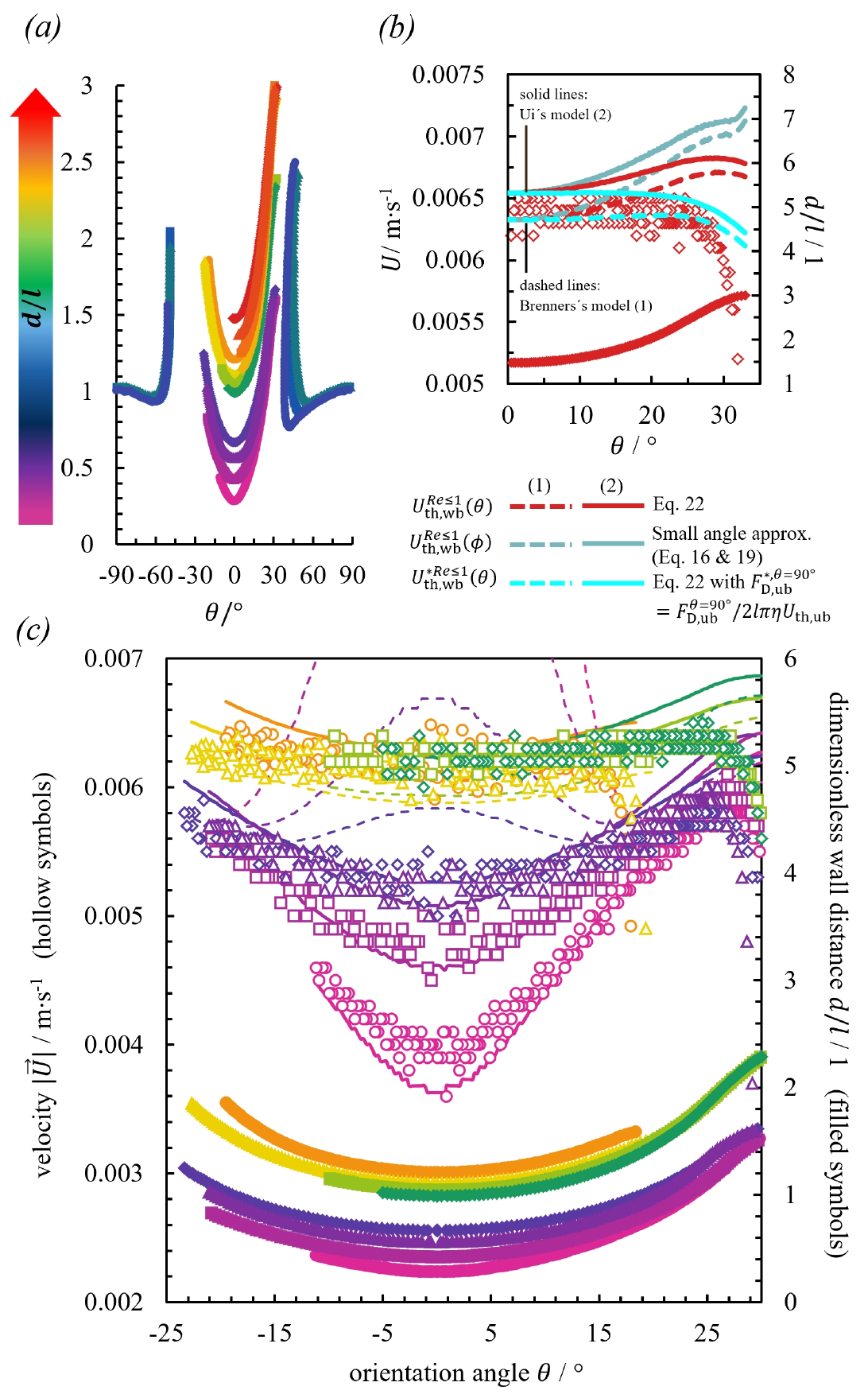}
    \caption{Translational kinematics of cylinders with a mean aspect ratio of
    $\bar{\chi}\approx5.8$ during near-wall sedimentation and comparison with
    the theoretical models. The cylinders were released at different initial
    wall distances and with different initial orientation angles.
    \textbf{(a)} Dimensionless wall distance $d/l$ plotted against the
    orientation angle $\theta$. The combined spatial and angular trajectories
    reveal glancing and contact-induced flipping behavior depending on the
    initial conditions.
    \textbf{(b)} Experimental translational velocity magnitude $|\vec{U}|$
    (hollow diamonds) and dimensionless wall distance $d/l$ (filled diamonds)
    plotted against $\theta$ for a cylinder released at a larger initial
    wall distance. The experimental velocity is compared with the theoretical
    velocity from Eq.~\eqref{eq22}, using the inertia-corrected axial and
    transverse drag expressions of Eqs.~\eqref{eq5} and \eqref{eq6}, in
    combination with Brenner's model (1) (dashed red line) and Ui's model (2)
    (solid red line). The green-gray lines show the small-angle
    approximations based on Ui's model (2), Eq.~\eqref{eq16}, and Brenner's
    model (1), Eq.~\eqref{eq19}, respectively. The red lines represent the
    model using a consistent normalization of both drag components, whereas
    the cyan line shows the corresponding result when the transverse drag
    component is normalized by $2\pi\eta lU_{\mathrm{th,ub}}$ instead of
    $4\pi\eta lU_{\mathrm{th,ub}}$.
    \textbf{(c)} Total translational velocity magnitude $|\vec{U}|$ (hollow
    symbols) and dimensionless wall distance $d/l$ (filled symbols) plotted
    against $\theta$ for cylinders released in the near-wall region
    ($d_{\mathrm{ini}}/l\lesssim2.5$). The experimental velocities are compared
    with Eq.~\eqref{eq22}, using Eqs.~\eqref{eq5} and \eqref{eq6} in
    combination with Ui's model (2) (solid lines) and Brenner's model (1)
    (dashed lines). The individual aspect ratios and initial wall distances
    and orientation angles corresponding to the different symbols are given
    in Fig.~\ref{Fig5}.}
    \label{Fig4}
\end{figure}

After establishing the sedimentation behavior in the symmetric configuration
at the center of the duct, the translational kinematics were investigated for
initially inclined cylinders with a mean aspect ratio of
\(\bar{\chi}\approx5.8\) released at different distances from a lateral wall.
The resulting asymmetric wall interaction leads to a coupled evolution of the
orientation angle \(\theta\), the dimensionless wall distance \(d/l\), and the
total translational velocity magnitude \(|\vec{U}|\) in the \(xy\)-plane.
Figure~\ref{Fig4} summarizes this evolution for the investigated initial
conditions and compares the measured velocities with the wall-corrected
theoretical models introduced in Sec.~\ref{sec2_1}. The individual aspect
ratios and initial conditions corresponding to the different experiments are
specified in Fig.~\ref{Fig5}.

As shown in Fig.~\ref{Fig4}(a), the near-wall trajectories depend strongly on
the initial wall distance and orientation. Cylinders released at larger wall
distances and moderate initial inclinations approach the wall while
simultaneously rotating toward the axially aligned orientation, i.e., toward
\(\theta=0^\circ\). For several trajectories, a minimum wall distance is
reached without contact, after which the cylinder moves away from the wall
while the orientation angle changes sign. This behavior qualitatively
corresponds to the \textit{glancing turn} described by Russel et al.
\cite{RusselWB.1977}, for which a sedimenting rod approaches a vertical wall,
rotates through the vertical orientation, and subsequently moves away with the
same end leading.

A distinctly different behavior is observed for cylinders released at
comparable initial wall distances but substantially larger initial orientation
angles. The three corresponding trajectories in Fig.~\ref{Fig4}(a) were
obtained for \(\theta_{\mathrm{ini}}\approx46.5^\circ\),
\(47.1^\circ\), and \(48.6^\circ\), at
\(d_{\mathrm{ini}}/l\approx2.50\), \(2.27\), and \(2.42\), respectively.
In contrast to the glancing trajectories obtained at similar initial wall
distances but smaller \(\theta_{\mathrm{ini}}\), these cylinders continue to
approach the wall until physical contact occurs. The subsequent interaction
with the wall induces a pronounced rotation through the horizontally aligned
orientation and a change in the sign of \(\theta\). This motion is referred
to here as \textit{contact-induced flipping behavior}. The comparison
demonstrates that the occurrence of glancing or flipping cannot be determined
from the initial wall distance alone and, for otherwise comparable wall
distances, is strongly dependent on the initial orientation.

The contact-induced flipping observed here should be distinguished from the
\textit{reversing turn} described by Russel et al.~\cite{RusselWB.1977}. In
their analysis of very slender rods, reversing denotes a non-contact motion at
large initial inclination, in which the rod changes its direction of rotation
and subsequently moves away from the wall with the opposite end leading.
Russel et al. additionally identified an intermediate range of initial
orientations for which the rod approaches sufficiently close to the wall that
collision occurs. The present flipping trajectories are therefore
qualitatively more closely related to this collision regime than to the
classical non-contact reversing turn. A direct comparison of the corresponding
transition angles is, however, not appropriate because the cylinders
investigated here have substantially smaller aspect ratios than the slender
rods considered by Russel et al.

The experiment shown in Fig.~\ref{Fig4}(b) exhibits the glancing turn
farthest from the wall, reaching $\theta=0^\circ$ at the largest wall
distance among the investigated trajectories. Figure~\ref{Fig4}(b)
compares the measured total translational velocity magnitude with the
predictions obtained using the different wall corrections. At larger
wall distances, particularly for $d/l\gtrsim2$, Brenner's model (1,
dashed lines) provides a closer description of the experimental
velocities than Ui's model (2, solid lines). As the cylinder approaches
the wall, however, the relative agreement changes, and Ui's model
provides a closer description of the measured velocities. Thus, neither
wall correction provides the closest agreement over the entire glancing
trajectory; their relative performance depends on the instantaneous
wall distance.

The corresponding small-angle approximations reproduce the measured
velocities well only in the immediate vicinity of the axially aligned
orientation, approximately within $|\theta|\lesssim5^\circ$. Outside
this restricted angular range, the predictions increasingly deviate
from the experimental data, consistent with the small-angle assumptions
underlying Eqs.~\eqref{eq16} and \eqref{eq19}.

The comparison in Fig.~\ref{Fig4}(b) also illustrates the influence of
the normalization of the transverse drag coefficient on the predicted
velocities. With the consistent normalization described in
Sec.~\ref{sec2_1}, in which the transverse drag coefficient is
normalized by $4\pi\eta l U_{\mathrm{th,ub}}$, the predictions based
on both wall corrections exhibit larger deviations from the
experimental data at large wall distances and large orientation
angles. For comparison, the cyan curve shows the result obtained when
the transverse drag coefficient is instead normalized by
$2\pi\eta l U_{\mathrm{th,ub}}$, while retaining the corresponding
velocity components in Eq.~\eqref{eq22}. This alternative normalization
leads to a closer agreement with the measurements in these regions.
However, it is not consistent with the normalization of the
translational drag coefficients adopted in Sec.~\ref{sec2_1} and
therefore does not provide a physically justified alternative to the
model based on the consistent normalization. The comparison
nevertheless highlights the limitations of the present wall-corrected
models at large wall distances and orientation angles.

Figure~\ref{Fig4}(c) focuses on the glancing trajectories with initial wall
distances $d_{\mathrm{ini}}/l\lesssim2.5$. A characteristic feature of these
experiments is the pronounced deceleration as the cylinders enter the
immediate vicinity of the wall, where $d/l<1$. The measured translational
velocities decrease sharply with decreasing wall distance, resulting in
strongly unsteady kinematics during this part of the trajectory. Despite
this pronounced unsteadiness, the measured velocities are well described
by the theoretical predictions based on Ui's wall-distance model (2, solid
lines), combined with the inertia-corrected axial and transverse drag
expressions. The agreement remains good even for $d/l<1$, where the
strongest deceleration is observed. In contrast, the prediction based on Brenner's model (1, dashed lines)
diverges as the wall distance decreases below $d/l=1$, yielding
unphysical velocity values at the smallest wall distances considered.

The quantitative comparison with the theoretical models is restricted to
trajectories exhibiting glancing behavior. For the contact-induced flipping
trajectories, substantial deviations between the measured and predicted
velocities already develop before physical contact with the wall. This is
consistent with the construction of the translational model in
Eq.~\eqref{eq22}, which interpolates between the limiting wall-bounded
velocities for axial and transverse cylinder orientations. At intermediate
orientations around $|\theta|\approx45^\circ$, the axial and transverse
contributions enter the interpolation with comparable weights, whereas
neither limiting configuration provides a close representation of the
hydrodynamics of an inclined cylinder. Consequently, deviations associated
with both limiting approximations can contribute simultaneously, making
this angular range particularly sensitive to the underlying model
assumptions. Once physical contact with the wall occurs, the cylinder
motion is additionally governed by contact mechanics and the associated
forces and torques, which are not included in the present hydrodynamic
model. The contact-induced flipping trajectories are therefore used to
characterize the observed motion but are excluded from the quantitative
comparison with the theoretical velocity models.

Corresponding experiments were also performed with shorter cylinders
having a mean aspect ratio of $\bar{\chi}\approx3.8$, and the results are
presented in Fig.~\ref{FigA1}. The principal trends observed for the
$\bar{\chi}\approx5.8$ cylinders are also reproduced at the smaller aspect
ratio. In particular, Brenner's model (1) provides a closer description
at larger wall distances, whereas Ui's model (2) better captures the
translational velocities as the minimum wall distance reached during
glancing decreases. The pronounced near-wall deceleration is likewise
observed for the shorter cylinders. However, the trajectory behavior
becomes less distinct at the smaller aspect ratio. In particular, the
distinction between glancing and contact-induced flipping is less
pronounced, with both types of motion occurring for comparable combinations
of initial wall distance and orientation. This increased sensitivity and
variability at smaller $\chi$ may be associated with a stronger influence
of finite-aspect-ratio effects, including the hydrodynamic influence of
the cylinder ends.

\subsection{Near-wall angular kinematics of initially inclined cylinders}
\label{sec4_3}

\begin{figure}
    \centering
    \includegraphics[width=\textwidth]{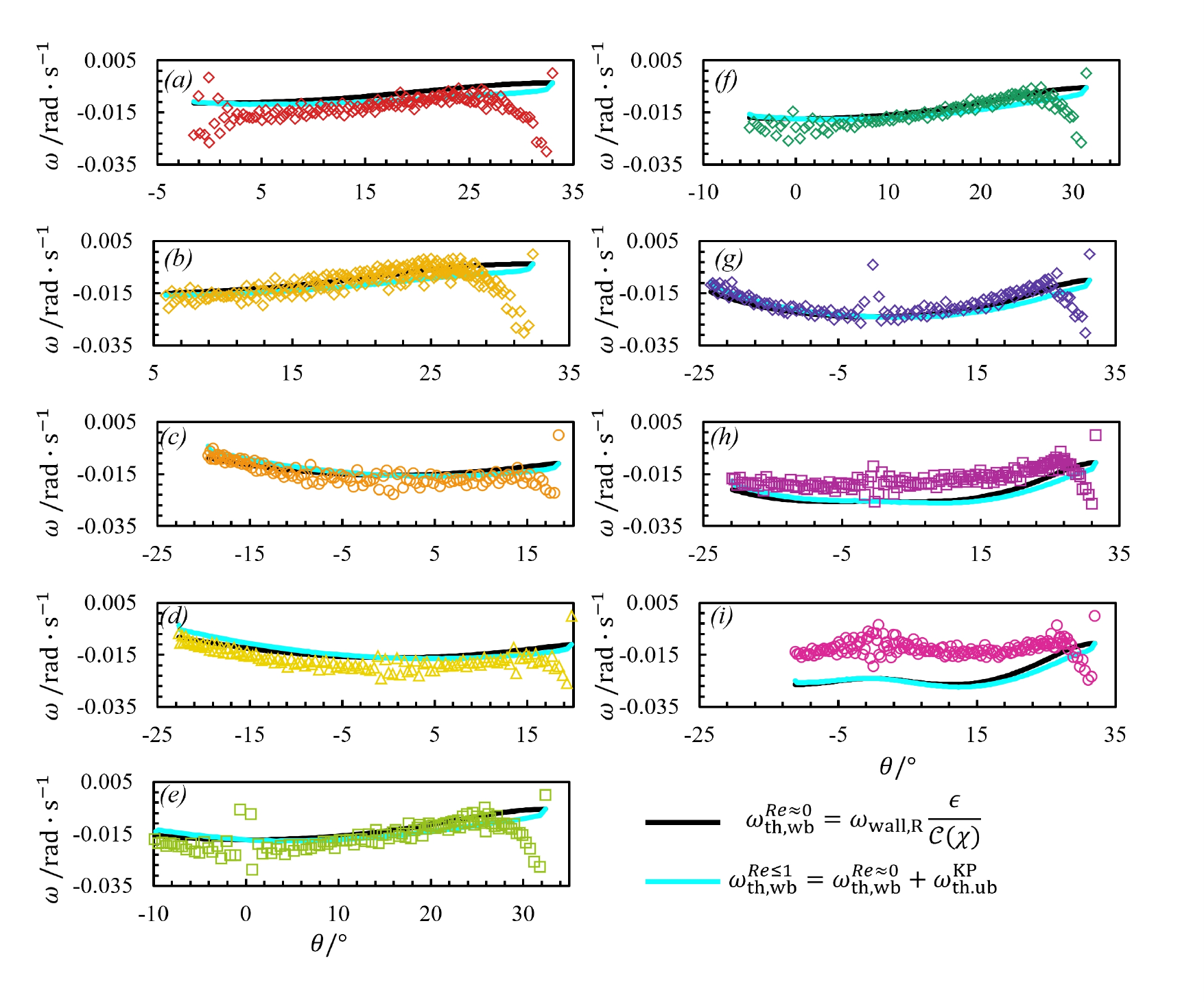}
    \caption{Angular kinematics of cylinders with a mean aspect ratio of
    \(\bar{\chi}\approx5.8\) during near-wall sedimentation and comparison
    with the theoretical models. Angular velocity \(\omega\) plotted against
    the orientation angle \(\theta\) for the individual experiments shown in
    Fig.~\ref{Fig4}. Hollow symbols represent the experimentally determined
    angular velocities. The black lines show the
    finite-aspect-ratio-corrected leading-order wall contribution based on
    Russel et al., Eq.~\eqref{eq37}, whereas the cyan lines show the
    combined angular-velocity model, Eq.~\eqref{eq40}, including the
    inertia-induced contribution from Eq.~\eqref{eq32}. The individually
    measured aspect ratios and initial conditions are:
    \textbf{(a)} \(\chi\approx6.0\), \(\theta_{\mathrm{ini}}\approx33.0^\circ\),
    \(d_{\mathrm{ini}}/l\approx3.00\);
    \textbf{(b)} \(\chi\approx6.3\), \(\theta_{\mathrm{ini}}\approx32.4^\circ\),
    \(d_{\mathrm{ini}}/l\approx2.91\);
    \textbf{(c)} \(\chi\approx5.7\), \(\theta_{\mathrm{ini}}\approx18.4^\circ\),
    \(d_{\mathrm{ini}}/l\approx1.59\);
    \textbf{(d)} \(\chi\approx6.0\), \(\theta_{\mathrm{ini}}\approx19.8^\circ\),
    \(d_{\mathrm{ini}}/l\approx1.51\);
    \textbf{(e)} \(\chi\approx5.9\), \(\theta_{\mathrm{ini}}\approx32.4^\circ\),
    \(d_{\mathrm{ini}}/l\approx2.39\);
    \textbf{(f)} \(\chi\approx6.3\), \(\theta_{\mathrm{ini}}\approx31.4^\circ\),
    \(d_{\mathrm{ini}}/l\approx2.34\);
    \textbf{(g)} \(\chi\approx5.8\), \(\theta_{\mathrm{ini}}\approx31.3^\circ\),
    \(d_{\mathrm{ini}}/l\approx1.67\);
    \textbf{(h)} \(\chi\approx6.0\), \(\theta_{\mathrm{ini}}\approx31.6^\circ\),
    \(d_{\mathrm{ini}}/l\approx1.59\); and
    \textbf{(i)} \(\chi\approx5.7\), \(\theta_{\mathrm{ini}}\approx32.0^\circ\),
    \(d_{\mathrm{ini}}/l\approx1.61\).}
    \label{Fig5}
\end{figure}

Following the analysis of the translational kinematics, the angular motion is
considered for the same near-wall sedimentation experiments shown in
Fig.~\ref{Fig4}. The analysis focuses on the glancing trajectories, for which
the continuous evolution of the angular velocity can be compared with the
theoretical models introduced in Sec.~\ref{sec2_2} without the additional
influence of physical wall contact. Figure~\ref{Fig5} shows the experimentally
determined angular velocity \(\omega\) as a function of the orientation angle
\(\theta\) for cylinders with a mean aspect ratio of
\(\bar{\chi}\approx5.8\). The individual panels correspond to the same
experiments presented in Fig.~\ref{Fig4}, with equivalent colors and symbols
to facilitate direct comparison with the corresponding \(d/l\)-\(\theta\)
trajectories. The experimental angular velocities are compared with the
finite-aspect-ratio-corrected leading-order wall contribution based on Russel
et al., Eq.~\eqref{eq37}, and with the combined angular-velocity model,
Eq.~\eqref{eq40}, which additionally includes the inertia-induced contribution
from Eq.~\eqref{eq32}.

Although the angular velocity is plotted as a function of \(\theta\) in
Fig.~\ref{Fig5}, the resulting curves should not be interpreted as a dependence
on the orientation angle alone. The wall-induced contribution in
Eq.~\eqref{eq37} is, to leading order, governed by the instantaneous
dimensionless wall distance \(d/l\), which itself evolves continuously with
\(\theta\) along each glancing trajectory. Consequently, the apparent
variation of \(\omega\) with \(\theta\) reflects primarily the coupled
evolution of the cylinder orientation and wall distance, \(\theta(t)\) and
\(d(t)/l\), rather than a purely angular dependence of the wall-induced
rotation. The corresponding \(d/l\)-\(\theta\) trajectories are shown in
Fig.~\ref{Fig4} and should therefore be considered together with the
angular-velocity curves in Fig.~\ref{Fig5}.

Before discussing the comparison with the theoretical models, two features of
Fig.~\ref{Fig5} should be noted. First, the panels are arranged according to
the wall distance reached at the axially aligned orientation. The dimensionless
wall distance at \(\theta=0^\circ\) progressively decreases from
Fig.~\ref{Fig5}(a) to (i). Thus, panel (a), corresponding to the experiment
shown in Fig.~\ref{Fig4}(b), represents the glancing trajectory passing through
the axially aligned orientation farthest from the wall, whereas the subsequent
panels represent trajectories reaching progressively smaller wall distances.
This arrangement facilitates a direct assessment of the increasing influence
of the wall on the angular kinematics.

Second, several experimental curves exhibit increased scatter in the measured
angular velocity in the immediate vicinity of \(\theta=0^\circ\). This scatter
is primarily an artifact of the image evaluation and numerical differentiation
rather than an indication of rapid physical fluctuations in the cylinder
rotation. The orientation angle is determined frame by frame from the segmented
cylinder geometry and is therefore subject to small variations associated with
the finite image resolution and segmentation. These comparatively small
uncertainties in \(\theta\) are amplified when calculating the angular velocity
from its temporal variation and become particularly apparent as the signed
orientation passes through \(\theta=0^\circ\). Accordingly, the individual
fluctuations in \(\omega\) close to the axially aligned orientation should not
be interpreted as physical unsteadiness of the rotational motion.

A common feature of the measured angular-velocity curves is the pronounced
maximum in \(|\omega|\) at the beginning of the evaluated trajectories,
corresponding to the largest positive orientation angles on the right-hand
side of the panels. As sedimentation proceeds, the magnitude of the angular
velocity initially decreases rapidly. This initial rotational transient
coincides with the translational acceleration phase following release. As
discussed in Sec.~\ref{sec3}, the measured initial orientation
\(\theta_{\mathrm{ini}}\) corresponds to the first orientation that can be
reliably determined after complete detachment from the vacuum pipette and
therefore differs from the preset release angle. During detachment, the
cylinder ends remain in close proximity to the pipette tips, while liquid
enters the developing gaps between the cylinder and the tips. The resulting
hydrodynamic interaction, together with the release process itself, can
therefore influence the initial translational and rotational motion.

Consequently, the pronounced initial variation in \(\omega\) cannot be
attributed unambiguously to an intrinsic transient hydrodynamic response of
the freely sedimenting cylinder. Such a contribution may in principle be
present, analogous to the transient inertial wall attraction previously
observed for spherical particles released from rest \cite{Noichl.2026}, but
cannot be separated from release-induced effects with the present
experimental configuration. The initial angular-velocity peak and its
subsequent rapid decrease are therefore reported here as part of the measured
kinematics but are not interpreted quantitatively in terms of the theoretical
near-wall rotation models.

A second common feature of the experiments is the good qualitative agreement
between the measured angular velocities and the theoretical predictions over
the entire range of investigated wall distances. In particular, the
theoretical curves reproduce the characteristic evolution of \(\omega\) along
the glancing trajectories, supporting the interpretation that the angular
kinematics are governed primarily by the changing wall distance. For the
moderate wall distances represented by Figs.~\ref{Fig5}(b)--(g), the agreement
is also quantitatively good over substantial parts of the trajectories. In
this range, the combined model, Eq.~\eqref{eq40}, differs only slightly from
the leading-order wall contribution, Eq.~\eqref{eq37}, indicating that the
additional inertia-induced contribution from Eq.~\eqref{eq32} remains
comparatively small.

A systematic difference between the two theoretical predictions nevertheless
becomes apparent when the approach toward and the motion away from the wall
are considered separately. During the approach toward the wall, i.e., before
the cylinder passes through the axially aligned orientation at
\(\theta=0^\circ\), inclusion of the inertia-induced contribution generally
shifts the prediction slightly away from the experimental data, such that
Eq.~\eqref{eq37} provides the closer description. After passage through
\(\theta=0^\circ\), when the cylinder moves away from the wall, the combined
model generally provides a slightly closer description of the measured
angular velocity. Thus, although the wall-induced contribution dominates the
predicted rotational motion, the influence of the additional inertia-induced
contribution depends on the respective part of the glancing trajectory.

More pronounced quantitative deviations are observed at the largest wall
distance in Fig.~\ref{Fig5}(a) and, in particular, for trajectories entering
the immediate wall region with \(d/l<1\). Nevertheless, even in the latter
regime, the theoretical curves reproduce the characteristic evolution of the
measured angular velocity remarkably well. This qualitative agreement
indicates that the leading-order wall-distance dependence retained in
Eq.~\eqref{eq37} captures an essential contribution to the rotational dynamics
even in the immediate wall region. The remaining quantitative deviations are
likely associated with several limitations of the theoretical description.

First, Eq.~\eqref{eq37} retains only the leading-order wall-distance dependence
based on the axially aligned configuration of Russel et al. and does not
explicitly include orientation-dependent contributions to the wall-induced
rotation. Its variation along a glancing trajectory therefore results
primarily from the changing instantaneous wall distance. At finite
inclination, particularly in the immediate vicinity of the wall, the
neglected orientation-dependent terms may become quantitatively important and
could explain why the model captures the characteristic evolution of
\(\omega\) while deviations remain in its absolute magnitude.

Second, higher-order contributions to the wall interaction are likewise not
represented by the leading-order description. Such contributions associated
with the wall-reflected disturbance flow are expected to become increasingly
relevant as the wall distance decreases. The good qualitative agreement
observed even for \(d/l<1\) therefore indicates that the leading wall-distance
dependence captures the dominant trend of the rotational motion, while
neglected higher-order wall contributions may account for part of the
remaining quantitative discrepancy.

Third, the strongest near-wall deviations occur in a region in which both the
translational and rotational kinematics vary substantially along the
trajectory. The theoretical wall contribution represents an instantaneous,
quasi-steady description of the hydrodynamic interaction and therefore
assumes that the surrounding disturbance flow adjusts sufficiently rapidly to
the changing particle configuration. At finite \(Re_{\mathrm P}\), a delayed
adjustment of the wall-reflected disturbance flow may contribute to the
observed deviations when the particle-wall configuration changes rapidly.
Such transient effects provide a plausible additional explanation for the
quantitative differences in the immediate wall region, although their
contribution cannot be isolated from the present experiments.

The quantitative discrepancy at the comparatively large wall distance in
Fig.~\ref{Fig5}(a) cannot readily be attributed to the same mechanisms that
become increasingly important in the immediate wall region. In this case, the
wall-induced rotational contribution is substantially weaker, and higher-order
near-wall corrections would not be expected to provide the primary
explanation. Deviations associated with the finite-aspect-ratio approximation,
the approximate representation of the weak wall interaction, and experimental
uncertainties may therefore become comparatively more apparent. The present
data do not permit the individual contributions to this far-field discrepancy
to be identified unambiguously.

The corresponding angular-velocity measurements for the shorter cylinders
with \(\bar{\chi}\approx3.8\) are shown in Fig.~\ref{FigA1}. The characteristic
evolution observed for the longer cylinders is reproduced, while the
quantitative agreement with the theoretical predictions is generally
improved. In particular, the model closely describes the measured angular
velocity for the glancing trajectory shown in Fig.~\ref{FigA1}(d). The same
trend was found for trajectories reaching even smaller wall distances,
including \(d/l<1\). Although the quantitative deviations increase in the
immediate wall region compared with those observed at intermediate wall
distances, they remain smaller than for the cylinders with
\(\bar{\chi}\approx5.8\) under comparable near-wall conditions. Thus, the
increasing deviations at \(d/l<1\) are observed at both aspect ratios, but
their magnitude depends on the cylinder aspect ratio. The improved agreement
for the shorter cylinders is also consistent with the aspect-ratio dependence
of the rotational-resistance expansion reported by Pierson et
al.~\cite{Pierson.2021}, although the respective contributions of the
resistance approximation and the approximations inherent in the wall model
cannot be separated from the present experiments.

Overall, the angular-velocity measurements demonstrate that the near-wall
rotation of the cylinders is governed predominantly by the wall-induced
contribution, with the instantaneous wall distance providing the principal
dependence along the glancing trajectories. The leading-order wall model
captures the characteristic evolution of the angular velocity over the entire
investigated range of wall distances and aspect ratios, including trajectories
entering \(d/l<1\), although its quantitative accuracy depends on both wall
distance and aspect ratio. The additional inertia-induced contribution remains
comparatively small under the present conditions and modifies the agreement
differently during the approach toward and departure from the wall. Together
with the translational results of Sec.~\ref{sec4_2}, these observations
demonstrate that the near-wall sedimentation of finite-aspect-ratio cylinders
is characterized by a strongly coupled evolution of wall distance,
orientation, translational velocity, and angular velocity, while comparatively
simple wall-corrected models already capture the principal features of the
experimentally observed kinematics.

\section{Summary and Conclusions}
\label{sec5}

The present study investigated the sedimentation of short, rigid cylinders in a rectangular duct filled with a highly viscous, initially quiescent liquid at particle Reynolds numbers of $Re_{\mathrm P}\approx O(10^{-1})$. Cylinders with mean aspect ratios of approximately $\bar{\chi}=3.8$--$5.8$ were released from rest, either approximately axially aligned near the centre of the duct or at different initial inclinations and distances from a lateral rigid wall. Time-resolved image measurements were used to determine their trajectories, orientations, translational velocities, and angular velocities. The measured kinematics were compared with theoretical descriptions accounting for finite-cylinder resistance, weak fluid inertia, and wall-induced hydrodynamic effects. The model velocities were evaluated using the instantaneously measured particle orientation and wall distance along the experimental trajectories; complete trajectories were not predicted from prescribed initial conditions.

For approximately axially aligned cylinders sedimenting near the centre of the rectangular duct, the experiments provided a reference configuration in which the influence of the surrounding walls could be assessed without the pronounced changes in orientation and lateral position observed in the near-wall experiments. The measured sedimentation velocities demonstrated that confinement remains relevant even when the cylinder is positioned approximately symmetrically between the duct walls. The comparison with the theoretical descriptions showed that accounting for weak fluid inertia improves the representation of the measured velocities and that the wall-correction approach based on Ui et al.~\cite{Ui.1984} provides closer agreement than the alternative based on Brenner~\cite{Brenner.1962} for the investigated centre-duct configurations. These results establish the importance of considering both finite-cylinder resistance and confinement when estimating the sedimentation velocity of short cylinders, even in a nominally symmetric configuration.

The experiments with initially inclined cylinders near a lateral wall revealed a coupling between translation and rotation. The evolving orientation and wall distance affected both the sedimentation velocity and the lateral motion, while wall-induced rotation altered the subsequent particle--wall configuration. Two distinct types of near-wall trajectories were observed. During glancing motion, the cylinder rotated through the axially aligned orientation in the vicinity of the wall and subsequently moved away without changing which end was leading. In contrast, contact-induced flipping involved physical contact with the wall and a subsequent rotation of the cylinder through a nearly horizontal orientation. This distinction is important because the non-contact reversing trajectories reported for very slender rods in earlier studies should not be identified with the contact-induced flipping observed here. The present velocity models do not include contact forces; consequently, their quantitative assessment was restricted to the glancing trajectories.

The translational measurements revealed different ranges of applicability for the investigated wall-correction models. At larger wall distances, the Brenner-based description generally provided a closer representation of the measured sedimentation velocities. At intermediate and small wall distances, the Ui-based description agreed more closely with the measurements, whereas the Brenner-based description failed in the close-wall regime. The translational velocity description incorporating both axial and transverse components reproduced the principal variations in the measured sedimentation velocities as the cylinders' orientations and wall distances evolved, although deviations remained in parts of the trajectories.

The angular measurements showed that the near-wall rotation of the investigated cylinders was governed predominantly by the wall-induced contribution represented by the leading-order term retained from the theory of Russel et al.~\cite{RusselWB.1977}. The inertia-induced contribution was comparatively small in the investigated Reynolds-number range. Including it did not uniformly improve agreement with the experiments: it tended to increase deviations during the approach toward the wall but improved the representation of parts of the subsequent departure. The angular-velocity model incorporating wall-induced and inertia-induced contributions reproduced the principal features of the measured rotational kinematics, including the change in angular velocity as the cylinder passed through the near-wall region. Quantitative agreement was generally closer at intermediate wall distances, whereas larger deviations occurred in some trajectories farther from the wall and in the immediate near-wall region. The latter discrepancies indicate that the leading-order wall description is insufficient to represent every stage of the motion, particularly when the cylinder approaches the boundary closely.

The experiments with the shorter cylinders, $\bar{\chi}\approx3.8$, were used to assess the applicability of the kinematic descriptions beyond the aspect ratios considered in the main near-wall experiments. The translational kinematics exhibited comparable trends, indicating that the principal features of the wall-induced changes in sedimentation velocity and lateral motion were also observed for the shorter cylinders. For the angular kinematics, the model showed even closer quantitative agreement with the measurements than for the longer cylinders. This finding supports the use of the higher-order slender-body approximation for rotational resistance in the present model: accounting for finite aspect ratio is important when extending a description originally developed for slender rods to short cylinders. However, the comparison alone does not establish that the higher-order approximation is the sole reason for the improved agreement, since the wall-induced contribution and the experimentally realized trajectories also affect the predicted angular velocity.

Taken together, the results show that the investigated velocity descriptions can reproduce important features of the instantaneous translational and angular kinematics of short cylinders sedimenting near a rigid wall, provided that their ranges of applicability are respected. Evaluating the models at the measured orientation and wall distance allows their predictions to be compared throughout an evolving experimental trajectory. Agreement with the measured instantaneous velocities does not, however, establish that the models can independently predict complete trajectories from initial conditions. The experiments also identify limitations associated with close wall proximity, larger inclinations, and contact-induced motion, for which additional hydrodynamic contributions or contact mechanics would need to be considered.

The present experiments provide a reference for investigating the near-wall sedimentation of short cylinders at lower particle Reynolds numbers and with additional influencing factors such as elasticity. Earlier experiments reported in the author's dissertation~\cite{Noichl.2026b} revealed markedly different kinematics at $Re_{\mathrm P}\approx O(10^{-2})$, including flipping of initially axially aligned cylinders and substantial variability between repeated releases. Additional, as yet unpublished experiments with initially inclined cylinders have also revealed flipping without contact with the wall. This non-contact behavior contrasts with the contact-induced flipping observed in the present study and demonstrates that the range of near-wall motions extends beyond the trajectories examined here. However, the greater variability observed at lower Reynolds numbers, together with the limited number of experiments available so far, precludes quantitative conclusions about the conditions governing these different motions. Further systematic measurements are needed to establish their dependence on Reynolds number, aspect ratio, initial orientation, and wall distance, and to determine whether the instantaneous velocity descriptions assessed in the present study remain applicable under these conditions.

% \section*{Declarations}

% \begin{itemize}
% \item Conflict of interest/Competing interests (check journal-specific guidelines for which heading to use)
% \item Ethics approval and consent to participate
% \item Consent for publication
% \item Materials availability
% \end{itemize}

\begin{appendices}
\section{Additional results}\label{secA1}

\begin{figure}
    \centering
    \includegraphics[width=0.9\textwidth]{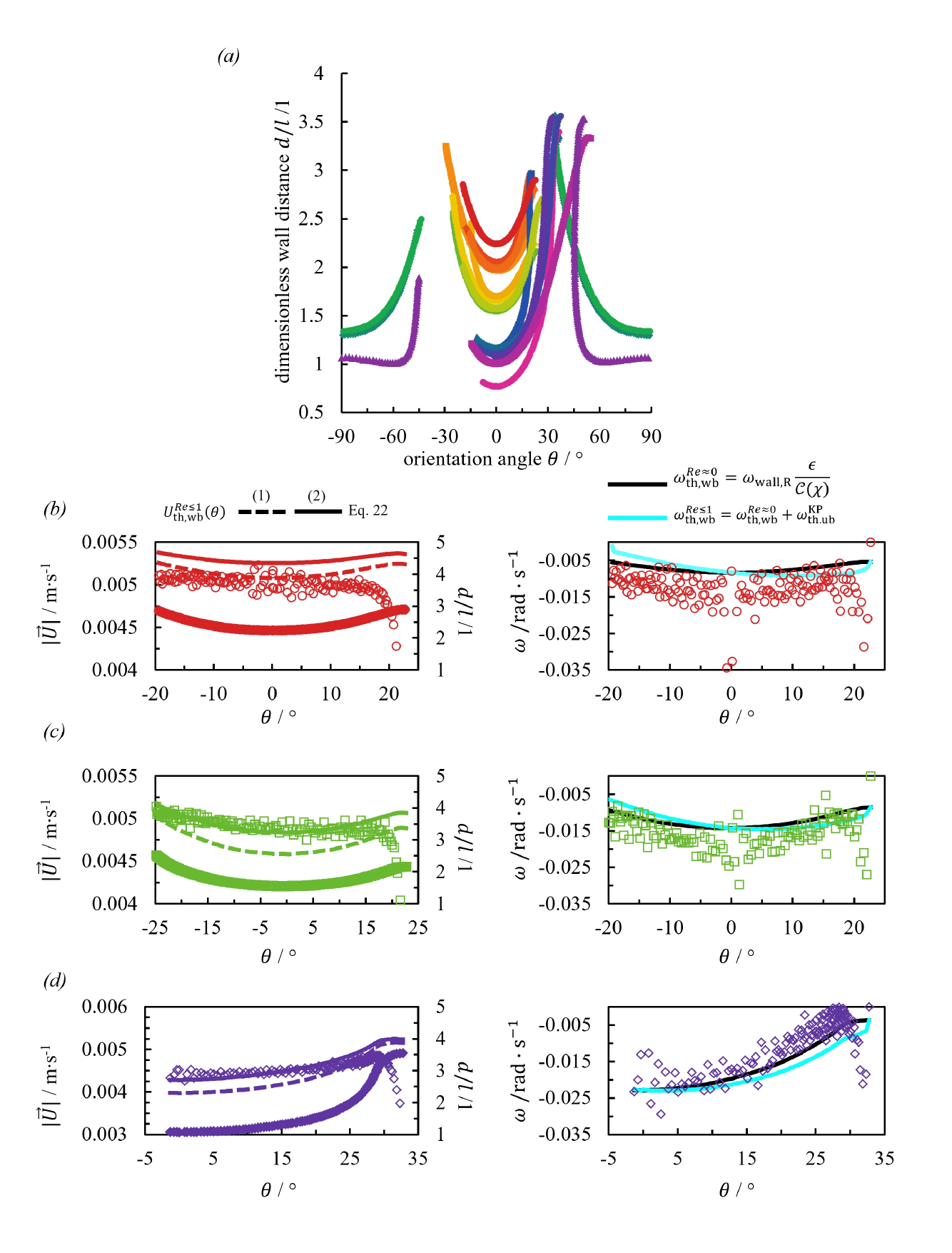}
    \caption{Translational and angular kinematics of shorter cylinders with a
    mean aspect ratio of \(\bar{\chi}\approx3.8\) during near-wall sedimentation
    and comparison with the theoretical models.
    \textbf{(a)} Dimensionless wall distance \(d/l\) plotted against the
    orientation angle \(\theta\) for cylinders released at different initial
    wall distances and orientation angles. The combined spatial and angular trajectories reveal glancing and contact-induced flipping behavior depending on the initial conditions.
    \textbf{(b--d)} Selected experiments exhibiting glancing behavior for three
    different initial conditions. The left panels show the total translational
    velocity magnitude in the \(xy\)-plane, \(|\vec{U}|\) (hollow symbols), and
    the dimensionless wall distance \(d/l\) (filled symbols) plotted against
    \(\theta\), together with the theoretical translational velocities based on Eq.~\eqref{eq22} with Brenner´s model (dashed lines) and Ui´s model solid lines. The right panels show the experimentally
    determined angular velocity \(\omega\) (hollow symbols) plotted against
    \(\theta\). The black lines represent the
    finite-aspect-ratio-corrected leading-order wall contribution based on
    Russel et al., Eq.~\eqref{eq37}, whereas the cyan lines represent
    the combined angular-velocity model, Eq.~\eqref{eq40}, including the
    inertia-induced contribution from Eq.~\eqref{eq32}, as described in
    Fig.~\ref{Fig5}. The individually measured aspect ratios and initial
    conditions of the selected experiments are:
    \textbf{(b)} \(\chi\approx3.8\), \(\theta_{\mathrm{ini}}\approx22.7^\circ\),
    \(d_{\mathrm{ini}}/l\approx2.90\);
    \textbf{(c)} \(\chi\approx3.9\), \(\theta_{\mathrm{ini}}\approx22.7^\circ\),
    \(d_{\mathrm{ini}}/l\approx2.16\); and
    \textbf{(d)} \(\chi\approx3.7\), \(\theta_{\mathrm{ini}}\approx32.7^\circ\),
    \(d_{\mathrm{ini}}/l\approx3.55\).}
    \label{FigA1}
\end{figure}

Figure~\ref{FigA1} shows the experimentally
determined dimensionless wall distance \(d/l\), total translational velocity magnitude \(|\vec{U}|\) and the angular velocity \(\omega\) as a function of the orientation angle
\(\theta\) for short cylinders with a mean aspect ratio of
\(\bar{\chi}\approx3.8\).

\end{appendices}

\bibliography{sn-bibliography}

\end{document}